\PassOptionsToPackage{dvipsnames}{xcolor}

\documentclass[sigconf, nonacm]{acmart}

\newcommand\vldbdoi{XX.XX/XXX.XX}

\newcommand\vldbavailabilityurl{https://github.com/brown-db/evergreen}
\newcommand\vldbpagestyle{plain} 

\usepackage{enumitem}
\usepackage{listings}
\usepackage{multirow}
\usepackage{subcaption}
\usepackage{tikz}
\usetikzlibrary{arrows.meta, positioning}

\newcommand{\baserm}{\texttt{base\_rm}} 
\newcommand{\error}{\varepsilon} 
\newcommand{\evgopt}{\texttt{evg\_opt}} 
\newcommand{\evgunopt}{\texttt{evg\_unopt}} 
\newcommand{\formula}{\varphi} 
\newcommand{\groups}{\mathcal{\group}} 
\newcommand{\group}{G} 
\newcommand{\interp}{\pi} 
\newcommand{\negtokens}{\overline{X}} 
\newcommand{\negtoken}{\overline{p}} 
\newcommand{\ngroups}{m} 
\newcommand{\ntuples}{n} 
\newcommand{\postokens}{X} 
\newcommand{\postoken}{p} 
\newcommand{\proportion}{\rho} 
\newcommand{\ragagent}{\texttt{rag\_agent}} 
\newcommand{\rank}{r} 
\newcommand{\relation}{\mathcal{R}} 
\newcommand{\rlm}{\texttt{rlm}} 
\newcommand{\sample}{S} 
\newcommand{\satset}{W} 
\newcommand{\system}{\textsc{Evergreen}} 
\newcommand{\tuple}{t} 

\begin{document}

\title{\system: Efficient Claim Verification for Semantic Aggregates}

\author{Alexander W. Lee}
\affiliation{%
  \institution{Brown University and Snowflake Inc.}
}
\email{alexander\_w\_lee@brown.edu}

\author{Benjamin Han}
\affiliation{%
  \institution{Snowflake Inc.}
}
\email{benjamin.han@snowflake.com}

\author{Shayak Sen}
\affiliation{%
  \institution{Snowflake Inc.}
}
\email{shayak.sen@snowflake.com}

\author{Sam Yeom}
\affiliation{%
  \institution{Snowflake Inc.}
}
\email{sam.yeom@snowflake.com}

\author{U\u{g}ur \c{C}etintemel}
\affiliation{%
  \institution{Brown University and Snowflake Inc.}
}
\email{ugur_cetintemel@brown.edu}

\author{Anupam Datta}
\affiliation{%
  \institution{Snowflake Inc.}
}
\email{anupam.datta@snowflake.com}

\begin{abstract}
With recent semantic query processing engines, semantic aggregation has become a primitive operator, enabling the reduction of a relation into a natural language aggregate using an LLM.
However, the resulting semantic aggregate may contain claims that are not grounded in the underlying relation.
Verifying such claims is challenging: they often involve quantifiers, groupings, and comparisons over relations that far exceed LLM context windows and require a costly combination of semantic and symbolic processing.

We present \system, a system that recasts claim verification as a semantic query processing task with tailored optimizations and provenance capture.
\system\ compiles each claim into a declarative semantic verification query that can execute on the same query engine used to produce the aggregate.
To reduce cost, \system\ avoids unnecessary LLM calls through verification-aware optimizations, including early stopping, relevance sorting, and estimation with confidence sequences, as well as general-purpose optimizations for semantic queries, such as operator fusion, similarity filtering, and prompt caching.
Each verdict is accompanied by citations that identify a minimal set of tuples justifying the result, with semantics based on semiring provenance for first-order logic.

On a benchmark of production-inspired workloads over restaurant review and customer support datasets, \system's optimized configurations occupy the entire cost--quality Pareto frontier.
With a strong LLM, \system\ preserves verification quality at an F1 of 0.94 while reducing cost by $3.1\times$ relative to unoptimized verification; with a substantially weaker LLM, it surpasses the strongest external baseline's F1 (0.87 vs.\ 0.83) at $7.0\times$ lower cost.
\end{abstract}

\maketitle

\pagestyle{\vldbpagestyle}

\ifdefempty{\vldbavailabilityurl}{}{
\vspace{.3cm}
\begingroup\small\noindent\raggedright\textbf{Artifact Availability:}\\
The source code, data, and/or other artifacts have been made available at \url{\vldbavailabilityurl}.
\endgroup
}

\begin{figure}[t]
  \centering
  \includegraphics[width=\columnwidth]{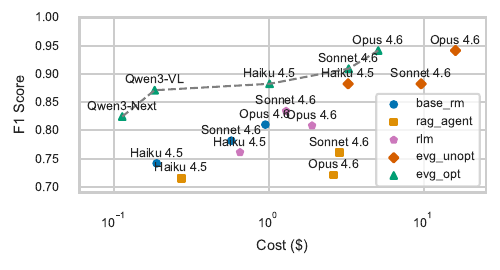}
  \caption{F1 vs.\ mean cost per claim (log scale), averaged over three trials. We compare \baserm\ (LLM-as-a-judge), \ragagent\ (retrieval-augmented agent), \rlm\ (recursive language model), \evgunopt\ (unoptimized \system), and \evgopt\ (optimized \system) across Claude and Qwen models. The dashed line connects the Pareto-optimal configurations. \evgopt\ exclusively occupies the Pareto frontier.}
  \Description{Scatter plot of F1 score vs.\ mean cost per claim for all configurations, showing that optimized \system\ exclusively occupies the Pareto frontier.}
  \label{fig:cost_f1_pareto}
\end{figure}

\section{Introduction}
\label{sec:introduction}

\emph{Semantic query processing engines} have recently emerged in both academia~\cite{urban_2023_caesura,patel_2025_lotus,liu_2025_palimpzest,lu_2025_vectraflow,anderson_2025_aryn,shankar_2025_docetl,chen_2025_continuous_prompts,sanmartino_2026_stretto,santos_2025_samsara,li_2025_docdb,wang_2025_aop,cheng_2023_binder} and industry~\cite{liskowski_2025_cortex_aisql,databricks_ai_functions,duckdb_prompt,bigquery_ai_functions}.
These systems feature \emph{semantic operators} that augment database operators with LLMs.
In several systems~\cite{patel_2025_lotus,shankar_2025_docetl,anderson_2025_aryn,liskowski_2025_cortex_aisql}, \emph{semantic aggregation} is a primitive operator, reducing a relation of tuples with an LLM-based aggregation function specified by a prompt.
For example, the prompt might summarize customer reviews or compare reviews across restaurants.
When the relation fits in the LLM's context window, the aggregation runs in a single LLM call; otherwise, it processes the relation in chunks and merges intermediate results.
Yet, the resulting \emph{semantic aggregate} may contain \emph{claims} that are not \emph{grounded} in the relation.
The LLM might hallucinate that ``the majority of reviews are positive'' when in fact only a minority are, or that ``no reviewers mention vegan options'' when one does.
The central question we investigate is: \textbf{how can we reliably and efficiently determine whether the claims embedded in a semantic aggregate are actually supported by the underlying relation?}
As semantic aggregation gains adoption in production systems such as Snowflake's Cortex AISQL~\cite{liskowski_2025_cortex_aisql}, answering this question becomes essential.

Verifying claims in semantic aggregates is challenging for several reasons.
First, the underlying relations are often large, easily exceeding LLM context windows; a relation of a few thousand reviews can span hundreds of thousands of tokens.
Second, from our observations of production workloads at Snowflake, claims that arise from such aggregations often involve quantifiers, groupings, and comparisons across groups (e.g., ``all locations have multiple complaints'', ``the top-ranked location is [A]'').
Verifying these claims requires semantic reasoning about each tuple followed by symbolic reasoning over the resulting values, a task LLMs struggle to accomplish reliably.
Third, semantically evaluating each tuple requires an expensive LLM invocation, making brute-force verification prohibitively costly.
Finally, a boolean verdict alone is often insufficient: users expect citations to the underlying data that explain it. 

Existing approaches to claim verification fall short for semantic aggregates.
Naive LLM-as-a-judge methods~\cite{min_2023_factscore,tang_2024_minicheck,zheng_2023_llm_judge} assess a claim against a provided source but assume that everything fits in the model's context window, an assumption violated by large relations.
Retrieval-augmented LLM approaches~\cite{khattab_2021_baleen,wang_2023_folk,xie_2025_fire,wei_2024_safe} scale to larger corpora by retrieving relevant evidence, but they target claims verified by connecting dependent sources rather than quantified claims over independent tuples, and their natural language reasoning about quantifiers and groupings is costlier and less reliable than dedicated symbolic operations.
Program-based LLM methods~\cite{zhang_2026_rlm,pan_2023_programfc,hu_2025_boost} can generate imperative programs that combine symbolic and semantic processing; however, they push the responsibility of optimization and citation capture onto the LLM, resulting in potentially suboptimal or error-prone implementations.
In the database community, query-based verification systems~\cite{jo_2019_aggchecker,karagiannis_2020_scrutinizer,jayasekara_2025_cedar,theologitis_2026_thucy,theologitis_2026_claimdb,chegini_2025_repanda} translate claims into declarative queries over structured data, but cannot handle claims that require semantic evaluation.
Meanwhile, semantic query processing engines are capable of expressing such verification logic.
Yet, existing systems lack verification-aware optimizations and explanations.
Most relevant to us is an early framework called \textsc{Binder}~\cite{cheng_2023_binder}, which naively processes all tuples with an LLM and was only evaluated on a claim verification benchmark limited to small structured tables~\cite{chen_2020_tabfact}.

We present \system, a system that reliably and efficiently verifies claims from semantic aggregates by treating claim verification as a semantic query processing problem with principled optimizations and provenance capture.
In particular, \system\ compiles each claim into a declarative \emph{semantic verification query} composed of standard relational operators and semantic operators, shifting the burden of optimization and citation capture away from the LLM and onto the query engine.
Crucially, \system\ can reuse the same query engine that produced the semantic aggregate, requiring no separate verification infrastructure.
Its efficiency stems from two complementary classes of optimizations.
First, \system\ introduces \emph{verification-aware optimizations}---including early stopping, relevance sorting, and estimation with anytime-valid confidence sequences~\cite{darling_1967_cs,waudby_smith_2023_betting_cs}---that exploit the structure of verification queries to minimize the amount of evidence that must be examined.
These techniques are specific to verification queries and constitute our primary technical contributions.
Second, \system\ incorporates \emph{general-purpose optimizations} for semantic queries, such as operator fusion, similarity filtering, and prompt caching.
Finally, \system\ accompanies each verdict with citations---a minimal explanation of why a claim holds or why not---formalized by applying and extending \emph{semiring provenance} for first-order logic~\cite{gradel_2024_fol_provenance}.

We evaluate \system\ on a benchmark of 32 diverse claims over three Yelp restaurant review datasets~\cite{yelp_open_dataset} and three Twitter customer support datasets~\cite{customer_support_on_twitter} based on production-inspired workloads.
Figure~\ref{fig:cost_f1_pareto} shows that across a range of LLMs, \system's optimized configurations form the entire cost--quality Pareto frontier.
Compared to an unoptimized version of \system\ with a strong LLM, its optimizations preserve verification quality at an F1 of 0.94 while reducing cost by 3.1$\times$.
The strongest external baseline~\cite{zhang_2026_rlm} reaches an F1 of 0.83, which \system\ surpasses at 0.87 with 7.0$\times$ lower cost using a substantially weaker LLM.
This robustness reflects \system's declarative, neuro-symbolic design: an LLM specifies only the high-level verification query, while symbolic query processing handles optimization and execution, restricting LLMs to fine-grained semantic tasks where they are reliable.

\smallskip
\noindent
\textbf{Summary of Contributions.}
\begin{itemize}[leftmargin=*]
  \item We identify several claim types that arise in semantic aggregates and map them to well-defined logical structures (Section~\ref{sec:claims}).
  \item We show how claims from semantic aggregates can be expressed as declarative semantic verification queries (Section~\ref{sec:queries}).
  \item We introduce optimizations tailored to verification and integrate them with general-purpose optimizations for semantic queries, substantially reducing cost while preserving quality (Section~\ref{sec:optimizations}).
  \item We build on semiring provenance for first-order logic~\cite{gradel_2024_fol_provenance} to formalize citations as data provenance, deriving a minimal explanation for each verdict (Section~\ref{sec:provenance}).
  \item We evaluate \system\ on a benchmark reflecting real-world semantic aggregation workloads, demonstrating its dominance in verification quality and cost over other approaches (Section~\ref{sec:evaluation}).
\end{itemize}

\section{Claims Over Relations}
\label{sec:claims}

Semantic aggregates implicitly encode a collection of \emph{claims}, i.e., natural language assertions about the underlying relation.
To verify these claims, \system\ first makes them explicit via a \emph{claim decomposition} process.
In particular, we follow prior work~\cite{wei_2024_safe} by first invoking an LLM to break down each sentence in the semantic aggregate into a list of claims.
Since the resulting claims may contain vague references (e.g., pronouns, unknown entities, non-full names), the process then invokes an LLM again for each claim to resolve the references using the original response as context.

From our analysis of claims generated by production semantic aggregation workloads at Snowflake, we identify several common claim types that map to familiar logical structures.
We first formalize our claim language and then describe each claim type.

Let $\relation$ be a \emph{relation}, which serves as the fixed, finite domain of our logic.
$\relation$ may be a filtered subset of the base relation relevant to the claim's scope (e.g., reviews for a specific restaurant or reviews from vegetarian customers).
That is, tuples not in $\relation$ are not relevant to the claim.
Let $\formula$ be a \emph{quantifier-free formula} with a free variable ranging over tuples $\tuple \in \relation$ (i.e., a unary query), mapping each tuple to true if it satisfies the formula (i.e., $\formula(\tuple) = \top$) or false if it does not (i.e., $\formula(\tuple) = \bot$).
$\formula$ may be built from symbolic predicates (e.g., whether the tuple's age attribute is greater than 21) and semantic predicates (e.g., whether the tuple's text attribute mentions vegan options) using conjunction, disjunction, and negation.

\textbf{Existential claims} use the existential quantifier to assert that \emph{at least one} tuple satisfies $\formula$, i.e., $\exists \tuple \in \relation : \formula(\tuple)$.
For example, ``Some reviewers enjoy the restaurant's chicken salad'' is existential.

\textbf{Universal claims} use the universal quantifier to assert that \emph{all} tuples satisfy $\formula$, i.e., $\forall \tuple \in \relation : \formula(\tuple)$.
For example, ``The reviews do not mention any vegan options'' is universal.

\textbf{Cardinal claims} use cardinal quantifiers to assert the \emph{number} of tuples that satisfy $\formula$, such as
$\exists_{\ge k} \tuple \in \relation : \formula(\tuple)$,
where $k \in \{0, 1, \ldots, |\relation|\}$.
For example, ``At least a handful of vegetarian customers enjoyed the restaurant's burgers'' is a cardinal claim.
Variants of cardinal quantifiers include $\exists_{> k}$, $\exists_{\le k}$, $\exists_{< k}$, $\exists_{= k}$, and $\exists_{\neq k}$.
Because $\relation$ is fixed and finite, cardinal quantifiers generalize existential and universal quantifiers with $\exists_{\ge 1}$ and $\exists_{= |\relation|}$, respectively.

\textbf{Proportional claims} use proportional quantifiers to assert the \emph{proportion} of tuples that satisfy $\formula$, such as
$\exists_{\ge \proportion} \tuple \in \relation : \formula(\tuple)$,
where $\proportion \in [0, 1]$.
For example, ``The restaurant has received majority positive reviews'' is a proportional claim.
Variants of proportional quantifiers include $\exists_{> \proportion}$, $\exists_{\le \proportion}$, $\exists_{< \proportion}$, $\exists_{= \proportion}$, and $\exists_{\neq \proportion}$.
Proportional quantifiers reduce to cardinal quantifiers by scaling $\proportion$ by $|\relation|$.

\textbf{Ordinal claims} assert that a group of tuples achieves a particular rank with respect to a per-group aggregate.
Let $\group_1, \ldots, \group_{\ngroups} \subseteq \relation$ be groups of tuples induced by partitioning the tuples in $\relation$ by one or more attribute values, and let $f(\group_j, \formula)$ denote a per-group aggregate (i.e., the count or proportion of tuples in $\group_j$ satisfying $\formula$).
Given the set $V(\formula) := \{f(\group_1, \formula), \ldots, f(\group_{\ngroups}, \formula)\}$ of distinct per-group aggregates, the rank of group $\group_i$ is
\[
\operatorname{rank}(\group_i, \formula) := |\{v \in V(\formula) : v > f(\group_i, \formula)\}| + 1.
\]
An ordinal claim asserts $\operatorname{rank}(\group_i, \formula) = \rank$ for some target group $\group_i$ and rank $\rank$.
For example, ``The top-ranked McDonald's location in terms of service has the Business ID [A]'' is an ordinal claim.

\textbf{Nested claims} compose the prior claim types at multiple depths of grouping.
Let $\mathcal{Q} := \{\exists, \forall, \exists_{\theta k}, \exists_{\theta \proportion}\}$ denote the full set of quantifiers introduced earlier with $\theta \in \{\ge, >, \le, <, =, \ne\}$.
Given a partition of groups $\groups := \{\group_1, \ldots, \group_{\ngroups}\}$ as before, a nested claim of depth two specifies an outer quantifier $Q_{\text{out}} \in \mathcal{Q}$ over groups $\group \in \groups$ and an inner quantifier $Q_{\text{in}} \in \mathcal{Q}$ over tuples $\tuple \in \group$:
\[
Q_{\text{out}} \group \in \groups : (Q_{\text{in}} \tuple \in \group : \formula(\tuple)).
\]
For example, ``All McDonald's locations have multiple complaints about poor service quality'' is $\forall \group \in \groups : (\exists_{\ge 2} \tuple \in \group : \formula(\tuple))$, where $\group$ is a group of tuples for a location and $\formula$ is the formula for whether a tuple mentions poor service quality.
Nesting can extend to arbitrary depths, though deeper nesting is rare in practice.

\section{Claims as Queries}
\label{sec:queries}

Based on the logical structure of claims, \system\ uses an LLM to compile natural language claims into \emph{semantic verification queries} composed of standard relational operators and semantic operators.
Akin to traditional database query languages, these verification queries are \emph{declarative}, enabling the LLM to focus on specifying \emph{what} to verify rather than \emph{how} to verify it.
This shifts low-level responsibilities such as optimization and provenance capture away from the LLM and onto the query engine, improving reliability and efficiency.
Moreover, verification requires no separate infrastructure: the same query engine that produced the semantic aggregate can also verify claims extracted from it.

\noindent
\textbf{Query Interface.}
\system\ provides a \texttt{DataFrame} API in Python, where a \texttt{DataFrame} \texttt{df} represents a relation $\relation$.
Queries are composed by chaining operators parameterized by expressions and then optimized and executed using \texttt{collect()}.
We detail the core operators and their relevant expressions, highlighting how they correspond to the logical structure of claims.

\textbf{\texttt{filter(predicate)}} selects tuples that satisfy the boolean expression \texttt{predicate}.
In verification queries, filters are generally used to restrict $\relation$ to tuples relevant to the claim.
In addition to symbolic expressions for structured filtering, \system\ features prompt expressions for semantic filtering: \texttt{prompt(prompt\_str, return\_type=bool)}.
The prompt expression uses an LLM to evaluate a semantic predicate specified by a prompt string \texttt{prompt\_str} that can reference tuple attributes by name.
\begin{lstlisting}
df.filter(prompt("The {text} mentions the restaurant's service"))
\end{lstlisting}

\textbf{\texttt{map(expr)}} computes a new attribute for each tuple by evaluating \texttt{expr} to materialize attributes required by the claim's formula $\formula$ that are not yet present in the relation.
As with filters, \texttt{expr} may be a symbolic or prompt expression, though prompt expressions here can also return non-boolean types.
\begin{lstlisting}
df.map(
  prompt("Identify the {text}'s sentiment", Sentiment).alias("sent")
)
\end{lstlisting}

\textbf{\texttt{aggregate(agg\_exprs, group\_by=())}} aggregates over tuples by evaluating the sequence of aggregation function expressions \texttt{agg\_exprs}, optionally grouped by a sequence of grouping expressions \texttt{group\_by}.
Aggregation functions are closely related to a claim's quantifiers, and an aggregation function's expression \texttt{expr} corresponds to the claim's formula $\formula$.
\texttt{bool\_or(expr)} corresponds to the existential quantifier, while \texttt{bool\_and(expr)} corresponds to the universal quantifier.
\texttt{count\_if(expr)} evaluates to the number of tuples satisfying \texttt{expr}, and when the resulting aggregate is paired with a comparison expression in a subsequent operator, it corresponds to a cardinal quantifier.
Similarly, \texttt{proportion(expr)} evaluates to the proportion of tuples satisfying \texttt{expr} and corresponds to a proportional quantifier when paired with a comparison expression.
\texttt{expr} can simply be a reference \texttt{col(name)} to a boolean attribute or an expression over attributes that evaluates to a boolean.
\begin{lstlisting}
df.aggregate(
  [proportion(col("sent").eq(Sentiment.POSITIVE)).alias("pos_prop")]
)
\end{lstlisting}
Grouping is used for ordinal and nested claims to compute group aggregates before subsequent ranking or outer aggregation steps.

\textbf{\texttt{with\_rank(expr, descending=True)}} creates a dense rank attribute \texttt{col("rank")} for each tuple based on the expression \texttt{expr}, in \texttt{descending} order by default.
This operator is used for ordinal claims, determining each group's rank after per-group aggregation.

\textbf{\texttt{check(predicate)}} is the terminal operator of a verification query.
It evaluates the boolean expression \texttt{predicate} on the (typically aggregated) result and adds it as a new boolean attribute representing the claim's verdict.
The verdict is annotated with provenance information that explains why the verdict is true or false, which we describe more in Section~\ref{sec:provenance}.

\noindent
\textbf{End-to-End Examples.}
Figure~\ref{fig:nested_query} shows the verification query for the nested claim from Section~\ref{sec:claims}.
The query first uses a semantic \texttt{map()} to evaluate $\formula$ on each tuple, labeling whether a review complains about poor service.
The first \texttt{aggregate()} groups tuples by location (\texttt{business\_id}) and applies \texttt{count\_if()} to count complaints per group for the inner cardinal quantifier $\exists_{\ge 2}$.
The second \texttt{aggregate()} applies \texttt{bool\_and()} to verify that every group's complaint count meets the threshold, implementing the outer universal quantifier $\forall$.
Finally, \texttt{check()} evaluates the verdict.
Figure~\ref{fig:ordinal_query} shows a second example with an ordinal claim.
The query additionally uses a semantic \texttt{filter()} to restrict to service-related reviews and \texttt{with\_rank()} to determine each location's rank.

\begin{figure}[t]
\begin{lstlisting}
df.map(
  prompt(
    "Identify whether the {text} complains about "
    "the restaurant's poor service", bool
  ).alias("service_complaint")
)
.aggregate(
  [count_if(col("service_complaint")).alias("complaint_count")],
  group_by=[col("business_id")]
)
.aggregate(
  [bool_and(col("complaint_count") >= 2).alias("all_mul_complaints")]
)
.check(col("all_mul_complaints"))
\end{lstlisting}
\caption{Verification query for the nested claim ``All McDonald's locations have multiple complaints about poor service quality'' ($\forall \group \in \groups : (\exists_{\ge 2} \tuple \in \group : \formula(\tuple))$).}
\Description{Verification query for the nested claim ``All McDonald's locations have multiple complaints about poor service quality''.}
\label{fig:nested_query}
\end{figure}

\begin{figure}[t]
\begin{lstlisting}
df.filter(prompt("The {text} mentions the restaurant's service"))
.map(
  prompt(
    "Identify whether the {text} praises the restaurant's service",
    bool
  ).alias("praises_service")
)
.aggregate(
  [proportion(col("praises_service")).alias("service_praise_prop")],
  group_by=[col("business_id")]
)
.with_rank(col("service_praise_prop"))
.filter(col("business_id").eq("[A]"))
.check(col("rank").eq(1))
\end{lstlisting}
\caption{Verification query for the ordinal claim ``The top-ranked McDonald's location in terms of service has the Business ID [A]'' ($\operatorname{rank}(\group_i, \formula) = 1$).}
\Description{Verification query for the ordinal claim ``The top-ranked McDonald's location in terms of service has the Business ID [A]''.}
\label{fig:ordinal_query}
\end{figure}

\section{Query Optimizations}
\label{sec:optimizations}

The bottleneck of semantic verification queries lies in their semantic operators, which require invoking expensive LLMs.
\system\ introduces verification-aware optimizations and combines them with general-purpose optimizations for semantic queries to reduce cost while maintaining reliability.

\subsection{Verification-Aware Optimizations}
\label{sec:verification-aware-optimizations}

\noindent
\textbf{Early Stopping.}
Computing full aggregations over all tuples is not required for many verification queries, since the aggregates in our setting are \emph{monotone} in the number of satisfying tuples.
As such, \system\ executes queries using the \emph{iterator model}---where operators lazily pull tuples from their children---and stops aggregating early so that only necessary tuples are pulled through the semantic operators that appear before the aggregation.
Table~\ref{tab:stopping-conditions} summarizes the conditions under which each aggregate stops deterministically and probabilistically, and we now describe how they are realized.

\texttt{bool\_or(expr)} stops aggregating after encountering a \emph{witness} (i.e., a tuple that satisfies \texttt{expr}), while \texttt{bool\_and(expr)} stops after encountering a \emph{counterexample} (i.e., a tuple that does not satisfy \texttt{expr}).
For \texttt{count\_if(expr)} and \texttt{proportion(expr)}, the query optimizer extracts the relevant comparison expressions from subsequent operators (e.g., \texttt{check(col("pos\_prop") > 0.5)}) and pushes them down to the corresponding aggregate operators as early stopping hints.
Given these hints, an accumulator stops as soon as the comparison is settled: either its running count $c$ already determines the outcome, or, when the total number $\ntuples$ of input tuples is known, the result cannot change regardless of whether the remaining tuples satisfy \texttt{expr}.
For an \emph{ungrouped} aggregate, $\ntuples$ is determined by a cardinality scan operator that counts the tuples prior to LLM evaluation.
While scanning is considered expensive in traditional query processing, it is effectively free when compared to LLM evaluation.
For a \emph{grouped} aggregate, $\ntuples$ is a per-group count computed during the sorting phase described next.

For grouped aggregates, \system\ avoids using hash-based aggregates since they require pulling all tuples through the semantic operators, which defeats the purpose of early stopping.
\system\ instead uses streaming aggregates, processing groups sequentially over sorted, group-contiguous input.
While streaming aggregates are classically only used when input tuples are \emph{already} sorted, sorting (similar to scans) is cheap in our setting.
The optimizer inserts a sort operator as low in the query plan as possible to ensure that tuples belonging to the same group are aggregated together while avoiding unnecessary semantic operator evaluations.
Once a group's aggregate is resolved, all operators below the aggregate skip the remaining tuples belonging to that group.

It is uncommon for traditional query engines to push comparison expressions into aggregates.
This optimization is, however, reminiscent of prior research in ad-hoc sensor networks that pushes \texttt{HAVING} predicates into \emph{grouped} monotonic aggregates to reduce transmission and storage of groups that do \emph{not} satisfy the predicate~\cite{madden_2003_tag}.
There, as in our setting, per-tuple processing is relatively expensive.
Yet, unlike this prior work, \system\ pushes these comparisons down to \emph{confirm} or \emph{refute} them as early as possible for both \emph{ungrouped} and \emph{grouped} aggregates.
We further amplify the benefits of this approach with the next two optimizations.

\noindent
\textbf{Relevance Sorting.}
Early stopping is most effective when relevant tuples---those that contribute to the aggregate's verdict---appear early in the input stream.
Since tuples are processed in arbitrary order by default, these relevant tuples may not surface until later.
\system\ encourages stopping by pre-sorting tuples by relevance.

At optimization time, the optimizer constructs a description of the aggregate to score each tuple's relevance.
It first traces the attribute referenced by the aggregate expression back to the semantic map that produced it, collecting the map's prompt expression and any filter prompts that reference the same text attribute.
These prompts provide context for what the aggregate is accumulating over (i.e., the scope of the claim's relation $\relation$ and the claim's formula $\formula$).
The description also includes the aggregate expression itself and, when applicable, its comparison expression; together, they express the claim's quantifier.
The optimizer then prompts an LLM with this description to generate a \emph{semantic search query}, a set of \emph{inclusion keywords} likely to appear in relevant tuples, and a set of \emph{exclusion keywords} that should not appear in them.
At execution time, each tuple is scored using three ranking signals: cosine similarity between the tuple's text embedding and the search query's embedding, the number of inclusion keywords present, and the number of exclusion keywords absent.
Each tuple's text embedding is computed during ingestion and reused across queries.
Signals are combined via Reciprocal Rank Fusion~\cite{cormack_2009_rrf} (with the standard constant $60$), and tuples are fed to the aggregate in descending relevance order.
The relevance sort operator is placed above the table scan so that reordering occurs before any LLM call.
This optimization is similar to a top-k operation in information retrieval, but allows the query engine to potentially process \emph{all} tuples in the relation without enforcing a set limit.
Again, computing relevance scores and sorting are inexpensive compared to LLM evaluation.

\begin{table}[t]
    \centering
    \scriptsize
    \caption{Deterministic and probabilistic stopping conditions to confirm/refute a claim. After observing $s$ of $\ntuples$ tuples with $c$ satisfying \texttt{expr}, the final count is in $[c, c + (\ntuples - s)]$. $[L_s, U_s]$ is the proportion CI and $\error$ the error tolerance. ``---'' marks a probabilistic condition subsumed by the deterministic one. $\exists_{> k}$ follows from $\exists_{\ge k+1}$. $\exists_{< k}$, $\exists_{\le k}$, and $\exists_{\ne k}$ swap confirm/refute of $\exists_{\ge k}$, $\exists_{> k}$, and $\exists_{= k}$, respectively. Proportional quantifiers replace counts with proportions and $k$ with $\proportion$.}
    \label{tab:stopping-conditions}
    \begin{tabular}{p{0.4cm} p{0.7cm} p{1.3cm} p{1.7cm} p{2.1cm}}
    \toprule
    & \multicolumn{2}{c}{Deterministic} & \multicolumn{2}{c}{Probabilistic} \\
    \cmidrule(lr){2-3}\cmidrule(lr){4-5}
    & Confirm & Refute & Confirm & Refute \\
    \midrule
    $\exists$ & $c \ge 1$ & $c + (\ntuples - s) < 1$ & --- & $\ntuples U_s < 1$ \\
    \addlinespace
    $\forall$ & $c = \ntuples$ & $c < s$ & $L_s \ge 1 - \error$ & --- \\
    \addlinespace
    $\exists_{\ge k}$ & $c \ge k$ & $c + (\ntuples - s) < k$ & $\ntuples L_s \ge k$ & $\ntuples U_s < k$ \\
    \addlinespace
    $\exists_{= k}$ & $c = k \land \newline s = \ntuples$ & $c > k \lor \newline c + (\ntuples - s) < k$ & $[\ntuples L_s, \ntuples U_s] \subseteq \newline [k(1 - \error), k(1 + \error)]$ & $[\ntuples L_s, \ntuples U_s] \cap \newline [k(1 - \error), k(1 + \error)] = \emptyset$ \\
    \bottomrule
    \end{tabular}
\end{table}

\begin{figure*}[t]
    \centering
    \scriptsize
    \begin{tikzpicture}[
      op/.style={draw, minimum width=1.5cm, minimum height=0.7cm, text width=1.5cm, align=center},
      ann/.style={text width=2cm, align=center},
      every path/.style={-{Latex}},
      node distance=0.6cm
    ]
    \node[op](scan){\texttt{table scan}};
    \node[op, right=of scan](rsort){\texttt{relevance sort}};
    \node[op, right=of rsort](sort){\texttt{sort} (\texttt{business\_id})};
    \node[op, right=of sort](shuf){\texttt{shuffle} (\texttt{business\_id})};
    \node[op, right=of shuf](map){\texttt{map}};
    \node[op, right=of map](inner){\texttt{count\_if} (\texttt{business\_id})};
    \node[op, right=of inner](outer){\texttt{bool\_and}};
    \node[op, right=of outer](check){\texttt{check}};
    
    \draw (scan) -- (rsort);
    \draw (rsort) -- (sort);
    \draw (sort) -- (shuf);
    \draw (shuf) -- (map);
    \draw (map) -- (inner);
    \draw (inner) -- (outer);
    \draw (outer) -- (check);
    
    \node[ann, below=0.1cm of rsort]{search query: ``...poor service...'', \\ incl: ``terrible'', ...};
    \node[ann, below=0.1cm of map]{prompt: \\ ``Identify whether the \texttt{\{text\}} complains...''};
    \node[ann, below=0.1cm of inner]{stop conditions: $c \ge 2$ or \\ $c + (\ntuples - s) < 2$};
    \node[ann, below=0.1cm of outer]{stop conditions: $L_s \ge 1 - \error$ or \\ $c = \ntuples$ or $c < s$};
    \end{tikzpicture}
    \caption{Optimized query plan for the verification query in Figure~\ref{fig:nested_query}.}
    \label{fig:nested_plan}
    \Description{Optimized query plan for the verification query in Figure~\ref{fig:nested_query}.}
\end{figure*}

\noindent
\textbf{Estimation with Confidence Sequences.}
In addition to deterministic early stopping, \system\ uses statistical estimation to resolve certain comparisons.
Our approach is similar to early work on online aggregation~\cite{hellerstein_1997_online_aggregation}, which uses classical confidence intervals (CIs).
However, using CIs for online aggregation suffers from the ``peeking problem''~\cite{johari_2017_peeking}, where continuously monitoring the CI inflates the error rate.
To avoid this issue, we instead leverage recent work on confidence sequences (CSs)~\cite{waudby_smith_2023_betting_cs}.

More formally, let $(Z_s)_{s \ge 1}$ be a sequence of Bernoulli random variables, where $Z_s = 1$ if the $s$-th tuple satisfies the aggregation function's boolean expression \texttt{expr} and $Z_s = 0$ otherwise.
Further let $\mu \in [0, 1]$ denote the true proportion of tuples satisfying \texttt{expr} in an accumulator's input stream.
The accumulator estimates $\mu$ with an \emph{anytime-valid $(1 - \alpha)$-confidence sequence}: a sequence of \emph{confidence intervals} $([L_s, U_s])_{s \ge 1}$ that satisfies
\[
\Pr(\forall s \ge 1 : \mu \in [L_s, U_s]) \ge 1 - \alpha,
\]
where $[L_s, U_s] \subseteq [0, 1]$ for all $s \ge 1$ and $\alpha \in (0, 1)$ is a configurable \emph{significance level}.
Equivalently, $1 - \alpha$ is the \emph{confidence level}.
Because the preceding guarantee holds simultaneously \emph{for all} sample sizes $s$, the accumulator can check $[L_s, U_s]$ after each observed tuple and stop as soon as the verdict is clear, without repeated checks inflating the error rate above $\alpha$.
A classical CI, by contrast, is valid only at a single, predetermined sample size, so continuously monitoring it is unsound.
CSs enable \system\ to stop confidently as early as possible without committing to a sample size that may be too small to be conclusive or too large to be cost-effective.

In general, a comparison resolves once the entire CI lies on one side of the threshold or within a tolerance interval (Table~\ref{tab:stopping-conditions}).
For \texttt{proportion()}, \system\ compares the proportion CI $[L_s, U_s]$ directly against $\proportion$; for \texttt{count\_if()}, it first scales the proportion CI by the total number $\ntuples$ of input tuples to obtain the count CI $[\ntuples L_s, \ntuples U_s]$ and compares against $k$.
Equality comparisons additionally use a configurable relative error tolerance $\error \in (0, 1)$.
For \texttt{bool\_and()}, the accumulator checks whether $\mu = 1$ using the equality condition with $\proportion = 1$, which reduces to $[L_s, U_s] \subseteq [1 - \error, 1]$, and therefore $L_s \ge 1 - \error$ since $U_s \le 1$ by construction.
For \texttt{bool\_or()}, since a witness already triggers deterministic stopping, the scaled count CI serves mainly to detect that no satisfying tuple exists, i.e., $\ntuples U_s < 1$.
When the total number $\ntuples$ of input tuples is known, we use sampling without replacement to yield tighter CIs.
\system\ computes CSs via the betting approach of~\citeauthor{waudby_smith_2023_betting_cs}~\cite{waudby_smith_2023_betting_cs}.

Since CSs require that observations arrive in exchangeable (i.e., random) order, the optimizer inserts a shuffle operator when estimation is used.
For ungrouped aggregates, the shuffle is placed above the table scan, randomizing all tuples.
For grouped aggregates, the shuffle is inserted above the sort operator used for grouping and hierarchically randomizes at each level: groups are shuffled as contiguous blocks, and tuples within the finest groups can also be shuffled if estimation is used at that aggregate level, preserving group contiguity for the streaming aggregate.

When multiple estimation decisions occur in the same query, \system\ controls the \emph{family-wise error rate}: the probability of \emph{any} CI failing to contain its true proportion must not exceed $\alpha$.
The error budget is divided in two stages.
First, for a query with $o$ aggregate operators that use estimation, the budget is split equally, allocating a significance level of $\alpha / o$ to each operator.
Second, within each operator, the budget is divided across its $a$ estimating accumulators and the $\ngroups$ groups in the input stream.
When the number of groups $\ngroups$ is known in advance from a sort operator, \system\ applies a Bonferroni correction, assigning each accumulator a per-group significance level of $\alpha / (o \cdot a \cdot \ngroups)$.
When $\ngroups$ is not known in advance, \system\ allocates a geometrically decreasing share to each successive group: each accumulator's $i$-th group receives a significance level of $\alpha / (o \cdot a \cdot 2^{i})$.
Because $\sum_{i=1}^{\infty} 1/2^{i} = 1$, the total error budget never exceeds $\alpha$ regardless of the number of groups.

\noindent
\textbf{Interaction Between Sorting and Estimation.}
Relevance sorting and estimation impose conflicting requirements on tuple order.
Relevance sorting arranges tuples to accelerate deterministic early stopping, while estimation requires exchangeable order for the CS guarantee.
The optimizer thus applies at most one strategy per aggregate operator.
The choice is guided by a heuristic that assumes claims are likely \emph{true}, which we expect to hold increasingly as LLMs improve.
Under this assumption, the deciding question is whether a few satisfying tuples suffice to confirm the claim: if so, relevance sorting surfaces them early; otherwise, the optimizer uses estimation.
Relevance sorting therefore applies to $\exists$, $\exists_{\ge k}$, and $\exists_{> k}$ claims with small $k$; estimation is unhelpful here, since tight CIs require observing more tuples.
The optimizer uses estimation for all other cases, where surfacing satisfying tuples early does not accelerate confirmation.
A true universal claim has $\mu = 1$, which estimation confirms from a sample without processing every tuple.
For $\exists_{\ge k}$ and $\exists_{> k}$ claims with large $k$, sorting does not pay off because too many satisfying tuples are needed.
$\exists_{\le k}$, $\exists_{< k}$, and $\exists_{= k}$ claims require bounding the total number of satisfying tuples, which entails processing nearly all tuples regardless of order.
A true $\exists_{\ne k}$ claim's count may lie above or below $k$; relevance sorting accelerates only the former by surfacing $k + 1$ satisfying tuples, whereas estimation can stop early in either case.
For proportional claims, the optimizer cannot determine whether the equivalent count threshold $\ntuples \proportion$ is small, since $\ntuples$ is unknown at optimization time.
In queries with nested aggregates, the two strategies can coexist at different levels: the inner aggregate uses relevance sorting within each group, while the outer aggregate uses estimation with group order shuffled.
We note that this bias toward true claims can backfire.
For example, if an existential claim is false, relevance sorting calls an LLM on the entire relation, while estimation would have been more efficient.
Adaptively selecting between relevance sorting and estimation at runtime is a promising future direction.

\noindent
\textbf{End-to-End Example.}
We illustrate how \system\ optimizes the query in Figure~\ref{fig:nested_query}, yielding the plan in Figure~\ref{fig:nested_plan}.
Before any LLM call, the plan reorders tuples: a relevance sort surfaces reviews likely to complain about poor service quality, and a sort on \texttt{business\_id} makes each location's reviews contiguous so \texttt{count\_if()}'s streaming aggregate can process one group at a time.
A shuffle then randomizes group order---preserving within-group relevance---so the outer \texttt{bool\_and()}'s CS observes groups in exchangeable order.
The semantic \texttt{map()} labels only the reviews pulled through.
The two aggregates apply different stopping conditions from Table~\ref{tab:stopping-conditions}: the inner \texttt{count\_if()} can only stop deterministically, while the outer \texttt{bool\_and()} can stop both deterministically and probabilistically.

\subsection{General-Purpose Optimizations}
\label{sec:general-purpose-optimizations}

\noindent
\textbf{Operator Fusion.}
A verification query can contain multiple semantic operators, e.g., a semantic filter followed by a semantic map, each requiring a separate LLM call per tuple.
\system\ reduces this cost by fusing consecutive operators, combining their prompts into a single LLM call.
If any filter evaluates to false, the fused operator discards the tuple; otherwise, it emits the map outputs.

When the fused operator contains a semantic filter, the cardinality scan operator---which provides an aggregate with the number $\ntuples$ of input tuples---is no longer used.
This is because neither placement of the cardinality scan works: placing it \emph{below} the fused operator yields a pre-filter count that \emph{overstates} the number of tuples the aggregate will observe, while placing it \emph{above} executes the fused operator's LLM on \emph{every} tuple before the aggregate begins (since it is a pipeline breaker), defeating the purpose of early stopping.
Without $\ntuples$, three capabilities described in Section~\ref{sec:verification-aware-optimizations} become unavailable: (1) deterministic bounds based on the maximum achievable count or proportion, (2) scaling the proportion CI to a count CI for \texttt{count\_if()}, and (3) tighter CIs via sampling without replacement.
In such cases, the optimizer falls back to other optimizations that do not require knowing $\ntuples$.
For instance, the inability to scale the CI prevents estimation for $\exists_{\ge k}$ and $\exists_{> k}$ cardinal claims, so the optimizer applies relevance sorting regardless of the threshold $k$.
Even though the stopping conditions that require $\ntuples$ are unavailable, operator fusion still reduces the total number of LLM calls.

Fusing multiple prompts into a single LLM call may reduce output quality.
Yet, based on our observations, verification queries usually require only a few narrowly scoped semantic operators.
Given the current capabilities of LLMs, fusion thus reduces the number of LLM calls with minimal quality degradation.

\noindent
\textbf{Similarity Filtering.}
Semantic filters invoke an LLM on every input tuple, even though many may not satisfy the predicate.
\system\ avoids these calls by inserting a similarity pre-filter below each semantic filter.
At optimization time, the optimizer reuses the same LLM-driven approach as in relevance sorting, generating a search query from the filter's prompt and embedding it.
At execution time, the pre-filter discards a tuple if the maximum cosine similarity between the query embedding and the tuple's sentence-level embeddings falls below a configurable threshold.
Sentence-level embeddings avoid the signal dilution of a single document embedding, retaining a tuple when even one of its sentences is relevant.
A conservatively low threshold keeps recall high while still eliminating tuples clearly unrelated to the predicate.

\noindent
\textbf{Prompt Caching.}
LLM responses are cached on disk, keyed by model name and prompt string.
When the same prompt is sent to a model, the cached response is reused to avoid redundant LLM evaluation.
This benefits workloads where multiple claims share semantic operators over the same underlying data.

\noindent
\textbf{End-to-End Example.}
The query in Figure~\ref{fig:ordinal_query} exercises the general-purpose optimizations.
A similarity pre-filter discards reviews unrelated to service before LLM evaluation, and the semantic \texttt{filter()} and \texttt{map()} are fused into a single LLM call per review.
Unlike the query in Figure~\ref{fig:nested_query}, ranking requires the full \texttt{proportion()} of each location, so the aggregate does not stop early; the savings instead come from issuing fewer LLM calls per tuple.
Finally, when related ordinal claims are verified (e.g., which location ranks first vs.\ second in service), prompt caching reuses LLM responses across claims.

\section{Citations as Provenance}
\label{sec:provenance}

Returning a boolean verdict is often not enough for users.
Each claim's verdict should be paired with a collection of citations so that users can check its underlying evidence.
By treating claims as queries, \system\ formalizes citations through the lens of \emph{data provenance}, giving users a precise explanation for the verdict based on facts in the underlying relation.
Specifically, we ground our work in \emph{semiring provenance for first-order logic}~\cite{gradel_2024_fol_provenance}.
We first provide a short overview of this formalization and then describe its application and extension to our setting.

\subsection{Semiring Provenance for First-Order Logic}

Seminal work on provenance semirings for \emph{positive relational algebra}~\cite{green_2007_provenance_semirings} annotates each output tuple with a \emph{polynomial} over a set of indeterminates $\postokens$ in a \emph{commutative semiring}.
Each \emph{indeterminate} $\postoken_{\tuple} \in \postokens$, also called a \emph{provenance token}, uniquely annotates an input tuple $\tuple \in \relation$.
The semiring's \emph{addition} ($+$) and \emph{multiplication} ($\cdot$) represent the \emph{alternative} (e.g., union) and \emph{joint} (e.g., join) use of input tuples, respectively.
Consequently, each \emph{monomial} in the polynomial is an \emph{alternative explanation} of which input tuples jointly contribute to the output tuple.
\citeauthor{gradel_2024_fol_provenance}~\cite{gradel_2024_fol_provenance} extend this framework to full \emph{first-order logic with negation} by pairing each indeterminate $\postoken_{\tuple} \in \postokens$ (a \emph{positive token}) with a \emph{dual indeterminate} $\negtoken_{\tuple} \in \negtokens$ (a \emph{negative token}).
In this semiring of \emph{dual-indeterminate polynomials}, addition corresponds to disjunction and existential quantification, while multiplication corresponds to conjunction and universal quantification.
In our setting, a positive token $\postoken_{\tuple}$ records the \emph{positive fact} that tuple $\tuple$ satisfies $\formula$ (i.e., $\formula(\tuple) = \top$), while a negative token $\negtoken_{\tuple}$ records the \emph{negative fact} that $\tuple$ does \emph{not} satisfy $\formula$ (i.e., $\formula(\tuple) = \bot$).
For example, given $\formula :=$ ``enjoys the restaurant's chicken salad'', $\postoken_{\tuple}$ records that tuple $\tuple$ enjoys the restaurant's chicken salad, while $\negtoken_{\tuple'}$ records that another tuple $\tuple'$ does \emph{not} enjoy it.

More formally, an \emph{interpretation} $\interp$ maps each literal to a polynomial and extends \emph{inductively} to all first-order formulas with the following rules (adapted from Definition 2 of~\cite{gradel_2024_fol_provenance}): (\romannumeral 1) $\interp\llbracket\psi \lor \chi\rrbracket := \interp\llbracket\psi\rrbracket + \interp\llbracket\chi\rrbracket$, (\romannumeral 2) $\interp\llbracket\psi \land \chi\rrbracket := \interp\llbracket\psi\rrbracket \cdot \interp\llbracket\chi\rrbracket$, (\romannumeral 3) $\interp\llbracket\exists \tuple \in \relation : \psi(\tuple)\rrbracket := \sum_{\tuple \in \relation} \interp\llbracket\psi(\tuple)\rrbracket$, (\romannumeral 4) $\interp\llbracket\forall \tuple \in \relation : \psi(\tuple)\rrbracket := \prod_{\tuple \in \relation} \interp\llbracket\psi(\tuple)\rrbracket$, and (\romannumeral 5) $\interp\llbracket\neg\psi\rrbracket := \interp\llbracket\operatorname{nnf}(\neg\psi)\rrbracket$,
where $\psi$ and $\chi$ are formulas and $\operatorname{nnf}(\cdot)$ denotes the transformation to negation normal form.
In our setting, we treat each $\formula(\tuple)$ and its negation $\neg\formula(\tuple)$ as literals regardless of $\formula$'s internal structure, so $\interp\llbracket\formula(\tuple)\rrbracket = \interp(\formula(\tuple))$ and $\interp\llbracket\neg\formula(\tuple)\rrbracket = \interp(\neg\formula(\tuple))$, which form the base cases for the inductive interpretation.
Since we want to track both positive and negative facts, we set $\interp(\formula(\tuple)) = \postoken_{\tuple}$ when $\formula(\tuple) = \top$ and $\interp(\formula(\tuple)) = 0$ otherwise; dually, $\interp(\neg\formula(\tuple)) = \negtoken_{\tuple}$ when $\formula(\tuple) = \bot$ and $\interp(\neg\formula(\tuple)) = 0$ otherwise.
That is, $0$ is interpreted as a \emph{false assertion}.

\citeauthor{gradel_2024_fol_provenance}'s main contribution~\cite{gradel_2024_fol_provenance} is support for \emph{negation}.
While the interpretation $\interp$ produces a polynomial explaining a \emph{true} sentence, it yields $0$ for a \emph{false} one, giving no explanation.
In this case, $\interp$ can instead be applied to the sentence's negation, which is true, revealing why the original is false.
This framework thus provides explanations for both cases, i.e., \emph{why-provenance} and \emph{why-not provenance}, so \system\ can justify each verdict, whether true or false.
We give examples of this approach in the next subsection.

\subsection{Provenance Semantics for Claims}
\label{sec:provenance-semantics}

\system\ derives its provenance from the prior interpretation rules.
However, rather than computing the full provenance polynomial (i.e., all alternative explanations), \system\ returns a single monomial (i.e., one explanation) either because the polynomial is itself a monomial or because of optimizations like early stopping.
By the semantics of the claim types, every complete monomial \system\ observes is \emph{minimal} (i.e., irredundant): no proper subset is also a monomial of the full polynomial.
This is immediate for true existential claims (a single token per monomial) and true universal claims (a single monomial over all tuples), and follows for the other claim types from their polynomials derived later in this section.
When the verdict is instead settled before a complete monomial is observed (e.g., estimation over a sample $\sample \subset \relation$ or deterministic early stopping from a maximum achievable count bound), the returned tokens form a \emph{subset} of such a minimal monomial.
Table~\ref{tab:provenance} summarizes the explanation \system\ returns for each claim type and verdict in complete monomial form, which we now derive.

\noindent
\textbf{Existential and Universal Claims.}
For a \emph{true existential} claim $\exists \tuple \in \relation : \formula(\tuple)$, applying rule (\romannumeral 3) yields $\sum_{\tuple \in \relation} \interp(\formula(\tuple))$.
Each tuple $\tuple$ contributes $\interp(\formula(\tuple)) = \postoken_{\tuple}$ to the polynomial if $\formula(\tuple) = \top$ and $\interp(\formula(\tuple)) = 0$ otherwise.
Since \system\ stops early and any monomial suffices as a \emph{witness}, the returned provenance is the corresponding positive token $\postoken_{\tuple}$ for the first tuple $\tuple$ where $\formula(\tuple) = \top$.
For example, given the existential claim ``Some reviewers enjoy the restaurant's chicken salad'', if $\tuple$ is the first tuple that satisfies $\formula :=$ ``enjoys the restaurant's chicken salad'', then the provenance is $\postoken_{\tuple}$.

For a \emph{false existential} claim, we first apply rule (\romannumeral 5) to interpret its \emph{negation} $\forall \tuple \in \relation : \neg\formula(\tuple)$. Applying rule (\romannumeral 4) then gives $\prod_{\tuple \in \relation} \interp(\neg\formula(\tuple)) = \prod_{\tuple \in \relation} \negtoken_{\tuple}$.
In this case, \system\ returns this single monomial containing all negative tokens, which indicates that no tuple in $\relation$ satisfies $\formula$ (e.g., no tuple enjoys the chicken salad).
When estimation is enabled and $\ntuples U_s < 1$, \system\ returns only $\prod_{\tuple \in \sample} \negtoken_{\tuple}$ for the processed sample $\sample \subset \relation$.

The universal cases are the dual of the existential cases.
For a \emph{true universal} claim $\forall \tuple \in \relation : \formula(\tuple)$, applying rule (\romannumeral 4) yields $\prod_{\tuple \in \relation} \interp(\formula(\tuple)) = \prod_{\tuple \in \relation} \postoken_{\tuple}$.
Hence, \system\ returns this monomial of all positive tokens, which indicates that each tuple in $\relation$ satisfies $\formula$.
When estimation is enabled and $L_s \ge 1 - \error$, \system\ returns only $\prod_{\tuple \in \sample} \postoken_{\tuple}$ for the processed sample.

For a \emph{false universal} claim, we apply rule (\romannumeral 5) to interpret its \emph{negation} $\exists \tuple \in \relation : \neg\formula(\tuple)$.
Applying rule (\romannumeral 3) then gives $\sum_{\tuple \in \relation} \interp(\neg\formula(\tuple))$.
Each tuple $\tuple$ contributes $\interp(\neg\formula(\tuple)) = \negtoken_{\tuple}$ if $\formula(\tuple) = \bot$ and $\interp(\neg\formula(\tuple)) = 0$ otherwise.
Since \system\ stops early and any monomial suffices as a \emph{counterexample}, the returned provenance is the corresponding negative token $\negtoken_{\tuple}$ for the first tuple $\tuple$ where $\formula(\tuple) = \bot$.

\begin{table}[t]
    \centering
    \scriptsize
    \caption{Explanation associated with each claim's verdict. $\satset \subseteq \relation$ is a set of tuples satisfying $\formula$. $\exists_{> k}$ follows from $\exists_{\ge k+1}$. $\exists_{< k}$, $\exists_{\le k}$, and $\exists_{\ne k}$ swap true/false explanations of $\exists_{\ge k}$, $\exists_{> k}$, and $\exists_{= k}$, respectively. Proportional claims reduce to the cardinal rows; ordinal and nested claims compose these rows.}
    \label{tab:provenance}
    \begin{tabular}{lll}
    \toprule
    & Explanation for True ($\top$) & Explanation for False ($\bot$) \\
    \midrule
    $\exists$ & $\postoken_{\tuple}$ (first witness) & $\prod_{\tuple \in \relation} \negtoken_{\tuple}$ \\
    $\forall$ & $\prod_{\tuple \in \relation} \postoken_{\tuple}$ & $\negtoken_{\tuple}$ (first counterexample) \\
    $\exists_{\ge k}$ & $\prod_{\tuple \in \satset} \postoken_{\tuple}$, $|\satset| = k$ & $\prod_{\tuple \in \satset} \postoken_{\tuple} \prod_{\tuple' \in \relation \setminus \satset} \negtoken_{\tuple'}$, $|\satset| < k$ \\
    $\exists_{= k}$  & $\prod_{\tuple \in \satset} \postoken_{\tuple} \prod_{\tuple' \in \relation \setminus \satset} \negtoken_{\tuple'}$, $|\satset| = k$ & $\prod_{\tuple \in \satset} \postoken_{\tuple} \prod_{\tuple' \in \relation \setminus \satset} \negtoken_{\tuple'}$, $|\satset| \ne k$ \\
    \bottomrule
    \end{tabular}
\end{table}

\noindent
\textbf{Cardinal and Proportional Claims.}
\citeauthor{gradel_2024_fol_provenance}~\cite{gradel_2024_fol_provenance} do not explicitly provide provenance expressions for cardinal quantifiers, so we derive them based on their interpretation rules.

First, $\exists_{\ge k} \tuple \in \relation : \psi(\tuple)$ is equivalent to $k$ nested existential quantifiers over strictly ordered tuples, each satisfying $\psi$.
Interpreting in the semiring replaces each existential quantifier with a sum and each conjunction with a product, yielding
\[
\interp\llbracket\exists_{\ge k} \tuple \in \relation : \psi(\tuple)\rrbracket = \sum_{\substack{\satset \subseteq \relation \\ |\satset|=k}} \prod_{\tuple \in \satset} \interp\llbracket\psi(\tuple)\rrbracket.
\]
Setting $\psi = \formula$, if $\formula(\tuple) = \top$ for all $\tuple \in \satset$, then $\prod_{\tuple \in \satset} \interp(\formula(\tuple)) = \prod_{\tuple \in \satset} \postoken_{\tuple} \neq 0$; otherwise, the product is $0$.
The polynomial is thus the sum of monomials, one for each size-$k$ combination of positive tokens.
The polynomial for $\exists_{> k}$ simply replaces $k$ with $k+1$.

In contrast, $\exists_{= k} \tuple \in \relation : \psi(\tuple)$ is equivalent to the same $k$ nested existential quantifiers, but additionally requires that every remaining tuple does \emph{not} satisfy $\psi$.
Interpreting in the semiring augments the $\exists_{\ge k}$ polynomial with a product over the remaining tuples $\tuple' \in \relation \setminus \satset$, each contributing $\interp\llbracket\neg\psi(\tuple')\rrbracket$, yielding
\[
\interp\llbracket\exists_{= k} \tuple \in \relation : \psi(\tuple)\rrbracket = \sum_{\substack{\satset \subseteq \relation \\ |\satset|=k}} \left( \prod_{\tuple \in \satset} \interp\llbracket\psi(\tuple)\rrbracket \prod_{\tuple' \in \relation \setminus \satset} \interp\llbracket\neg\psi(\tuple')\rrbracket \right).
\]
Setting $\psi = \formula$, if $\formula(\tuple) = \top$ for all $\tuple \in \satset$ and $\formula(\tuple') = \bot$ for all $\tuple' \in \relation \setminus \satset$, then we have that $\prod_{\tuple \in \satset} \interp(\formula(\tuple)) \prod_{\tuple' \in \relation \setminus \satset} \interp(\neg\formula(\tuple')) = \prod_{\tuple \in \satset} \postoken_{\tuple} \prod_{\tuple' \in \relation \setminus \satset} \negtoken_{\tuple'} \neq 0$; otherwise, the product is $0$.
Since each tuple either satisfies $\formula$ or does not, at most one subset $\satset$ yields a nonzero product, precisely when $|\satset| = k$, i.e., the sentence holds.
The polynomial is then a single monomial $\prod_{\tuple \in \satset} \postoken_{\tuple} \prod_{\tuple' \in \relation \setminus \satset} \negtoken_{\tuple'}$: a positive token for each of the $k$ tuples satisfying $\formula$ and a negative token for the remaining $|\relation| - k$.
The provenance polynomials for $\exists_{< k}$, $\exists_{\le k}$, and $\exists_{\neq k}$ are the same as $\exists_{= k}$ but sum over subsets $\satset$ of size $< k$, $\le k$, and $\neq k$, respectively.
As with $\exists_{= k}$, requiring every tuple outside $\satset$ not to satisfy $\formula$ forces $\satset$ to be exactly the set of satisfying tuples; every other subset contributes $0$.
Hence, the polynomials for the other cardinal quantifiers also reduce to at most the single monomial $\prod_{\tuple \in \satset} \postoken_{\tuple} \prod_{\tuple' \in \relation \setminus \satset} \negtoken_{\tuple'}$ where $\satset$ is the satisfying set.
The size constraint on $\satset$ ($< k$, $\le k$, or $\neq k$) only determines whether this monomial is included in the sum, i.e., whether the polynomial is nonzero and thus the sentence holds.

Given these polynomials, we now describe the provenance \system\ returns for each cardinal claim type.
For a \emph{true cardinal} claim $\exists_{\ge k} \tuple \in \relation : \formula(\tuple)$, at least one monomial $\prod_{\tuple \in \satset} \postoken_{\tuple}$ of $|\satset| = k$ positive tokens exists.
Since early stopping halts after finding $k$ satisfying tuples, \system\ returns only these first $k$ positive tokens.
For example, if the claim ``At least a handful of vegetarian customers enjoyed the restaurant's burgers'' is true, \system\ returns the first $k$ positive tokens with tuples satisfying $\formula :=$ ``vegetarian customer who enjoyed the restaurant's burgers''.
Similar to the prior claim types, when estimation is enabled and $\ntuples L_s \ge k$, \system\ returns only the positive tokens in the processed sample $\sample \subset \relation$.

For a \emph{false cardinal} claim $\exists_{\ge k} \tuple \in \relation : \formula(\tuple)$, its \emph{negation} $\exists_{< k} \tuple \in \relation : \formula(\tuple)$ is true, so its provenance is $\prod_{\tuple \in \satset} \postoken_{\tuple} \prod_{\tuple' \in \relation \setminus \satset} \negtoken_{\tuple'}$ where $|\satset| < k$, which demonstrates that fewer than $k$ tuples satisfy $\formula$.
When only a subset of tuples is processed---either because the count provably cannot reach $k$ given the remaining tuples, or because $\ntuples U_s < k$---the tokens cover only the processed tuples.

For $\exists_{= k}$ and $\exists_{\ne k}$, when the count equals $k$, \system\ processes all tuples and outputs all tokens to demonstrate the exact count; when the count differs from $k$, it may stop early and output the processed tokens, demonstrating that the count lies above or below $k$.
Since proportional claims reduce to cardinal claims, \system\ scales the proportion threshold $\proportion$ by the number of processed tuples and then applies the provenance semantics of cardinal claims.

\noindent
\textbf{Ordinal Claims.}
An ordinal claim asserts that a target group $\group_i$ achieves rank $\rank$ among $\ngroups$ groups, where the rank is based on a per-group aggregate $f$.
This is determined by a conjunction of $\ngroups - 1$ pairwise cardinal comparisons between $\group_i$ and every other group $\group_j$ ($i \ne j$); in the semiring, this conjunction becomes a product of per-comparison polynomials.
Each comparison produces two monomials (one per group) following the provenance semantics for cardinal or proportional claims depending on the aggregate $f$.
When $f(\group_i, \formula) > f(\group_j, \formula)$, $\group_i$'s tokens follow the rules for a true $\exists_{> f(\group_j, \formula)}$ claim, while $\group_j$'s tokens follow the rules for a true $\exists_{< f(\group_i, \formula)}$ claim.
When $f(\group_i, \formula) < f(\group_j, \formula)$, the roles are reversed.
When $f(\group_i, \formula) = f(\group_j, \formula)$, both groups' tokens follow the rules for a true $\exists_{= f(\group_i, \formula)}$ claim and contribute all tokens.

\noindent
\textbf{Nested Claims.}
A nested claim composes an outer quantifier over groups with an inner claim per group.
Such claims are already supported by the \emph{inductive} interpretation rules.
\system\ first evaluates each group's inner claim, producing per-group provenance tokens following the semantics of the inner claim type.
The outer quantifier then combines these per-group tokens following its provenance semantics.
For example, if the claim $\forall \group \in \groups : (\exists_{\ge 2} \tuple \in \group : \formula(\tuple))$ is true, the provenance is the product of two positive tokens per group.
Otherwise, the provenance is the first failing group's tokens, showing that fewer than two of its tuples satisfy $\formula$.

\section{Experimental Evaluation}
\label{sec:evaluation}

Our experimental evaluation aims to show that \system\ achieves higher verification quality at lower cost than existing approaches.
Beyond this end-to-end comparison, we analyze the quality of \system's provenance and semantic operators, and ablate the effect of each optimization on quality and cost.

\begin{table*}[t]
    \centering
    \caption{Benchmark claims.}
    \label{tab:claims}
    \scriptsize
    \begin{tabular}{lllllc}
      \toprule
      ID & Dataset & Semantic Agg. Type & Claim Type & Claim & Grounded \\
      \midrule
      C1 & \multirow{4}{*}{\texttt{johns}} & \multirow{4}{*}{Summarize} & $\exists_{< 5}$ & Less than a handful of customers complained about service quality. & $\bot$ \\
      C2 & & & $\exists$ & Some reviewers enjoy the restaurant's chicken salad. & $\top$ \\
      C3 & & & $\exists_{\ge 0.1}$ & Common criticisms of John's Roast Pork include cash-only policy. & $\bot$ \\
      C4 & & & $\forall$ & The reviews do not mention any vegan options. & $\top$ \\
      \addlinespace
      C5 & \multirow{8}{*}{\texttt{mcdonalds}} & \multirow{4}{*}{Compare} & $\exists \forall$ & Some McDonald's locations had no negative reviews. & $\bot$ \\
      C6 & & & $\exists_{< 0.5} \exists_{\ge 2}$ & Only a minority of McDonald's locations had multiple reports of incorrect orders. & $\bot$ \\
      C7 & & & $\exists_{> 0.5} \exists$ & The majority of McDonald's locations had reports of cold food. & $\top$ \\
      C8 & & & $\forall \exists_{\ge 2}$ & All McDonald's locations have multiple complaints about poor service quality. & $\top$ \\
      \addlinespace
      C9 & & \multirow{4}{*}{Rank} & $\operatorname{rank}() = 1$ & The top-ranked McDonald's location in terms of service has the Business ID [A]. & $\top$ \\
      C10 & & & $\operatorname{rank}() = 2$ & The McDonald's location ranked \#2 in terms of service has the Business ID [B]. & $\top$ \\
      C11 & & & $\operatorname{rank}() = 1$ & The top-ranked McDonald's location in terms of service has the Business ID [B]. & $\bot$ \\
      C12 & & & $\operatorname{rank}() = 2$ & The McDonald's location ranked \#2 in terms of service has the Business ID [A]. & $\bot$ \\
      \addlinespace
      C13 & \multirow{4}{*}{\texttt{village}} & \multirow{4}{*}{Summarize} & $\exists_{> 5}$ & More than a handful of vegetarian customers enjoyed the restaurant's burgers. & $\top$ \\
      C14 & & & $\exists$ & Some reviewers enjoyed the restaurant's calamari. & $\bot$ \\
      C15 & & & $\exists_{> 0.5}$ & Village Whiskey has received majority positive reviews. & $\top$ \\
      C16 & & & $\forall$ & There are over 200 varieties of whiskey available at Village Whiskey. & $\bot$ \\
      \addlinespace
      C17 & \multirow{4}{*}{\texttt{play}} & \multirow{4}{*}{Summarize} & $\exists_{= 240}$ & 240 customers are experiencing problems with their accounts or login issues. &  $\bot$ \\
      C18 & & & $\exists_{\ne 240}$ & The number of customers experiencing problems with their accounts or login issues is not 240. & $\top$ \\
      C19 & & & $\exists_{= 0.1}$ & 10\% of customers are experiencing problems with payments or billing. & $\bot$ \\
      C20 & & & $\exists_{\ne 0.1}$ & The percentage of customers experiencing problems with payments or billing is not 10\%. & $\top$ \\
      \addlinespace
      C21 & \multirow{8}{*}{\texttt{airlines}} & \multirow{4}{*}{Compare} & $\exists_{= 4} \exists_{> 0.15}$ & There are 4 airline companies where over 15\% of customer complaints are regarding flight delays. & $\top$ \\
      C22 & & & $\exists_{= 2} \forall$ & There are 2 airline companies where the support agent apologizes in all dialogs with a customer complaint. & $\bot$ \\
      C23 & & & $\exists \exists_{s^+ \ge 0.2 \wedge s^- \ge 0.2}$ & The customer satisfaction for some airlines was a mixed bag. & $\top$ \\
      C24 & & & $\forall \exists$ & All airlines have customer complaints about the taste of the food provided on their flight. & $\bot$ \\
      \addlinespace
      C25 & & \multirow{4}{*}{Rank} & $\operatorname{rank}() = 1$ & AirAsiaSupport has the most customer support dialogs with complaints about flight booking issues. & $\top$ \\
      C26 & & & $\operatorname{rank}() = 2$ & SouthwestAir has the second most customer support dialogs with complaints about flight booking issues. & $\top$ \\
      C27 & & & $\operatorname{rank}() = 1$ & SouthwestAir has the most customer support dialogs with complaints about flight booking issues. & $\bot$ \\
      C28 & & & $\operatorname{rank}() = 2$ & AirAsiaSupport has the second most customer support dialogs with complaints about flight booking issues. & $\bot$ \\
      \addlinespace
      C29 & \multirow{4}{*}{\texttt{uber}} & \multirow{4}{*}{Summarize} & $\exists_{= 776}$ & 776 customers complained about poor driver behavior. & $\top$ \\
      C30 & & & $\exists_{< 725}$ & Less than 725 customers complained about poor driver behavior. & $\bot$ \\
      C31 & & & $\exists_{< 0.1}$ & Less than 10\% of customers reported receiving inconsistent or inaccurate information from support agents. & $\bot$ \\
      C32 & & & $\exists_{< 0.25}$ & Less than 25\% of customers reported receiving inconsistent or inaccurate information from support agents. & $\top$ \\
      \bottomrule
    \end{tabular}
\end{table*}

\noindent
\textbf{Benchmark Datasets and Claims.}
To the best of our knowledge, no benchmark targets claim verification over semantic aggregates, so we curate one to evaluate \system.
From our analysis of production semantic aggregation queries on Snowflake AISQL~\cite{liskowski_2025_cortex_aisql}, we observe that queries are often executed over relations of customer reviews and support tickets.
As such, we first select three subsets of the restaurant reviews in the Yelp Open Dataset~\cite{yelp_open_dataset}, which we identify as \texttt{johns} (1,609 tuples; 230k tokens), \texttt{mcdonalds} (1,813 tuples; 244k tokens), and \texttt{village} (1,603 tuples; 291k tokens).
\texttt{mcdonalds} contains reviews from 62 McDonald's locations in Missouri, while \texttt{johns} and \texttt{village} are reviews of separate restaurants.
We then include three subsets of the Customer Support on Twitter dataset~\cite{customer_support_on_twitter}, which we call \texttt{play} (1,197 tuples; 394k tokens), \texttt{airlines} (3,167 tuples; 1,213k tokens), and \texttt{uber} (2,687 tuples; 1,005k tokens).
\texttt{airlines} contains dialogs between customers and support agents across six airlines, while \texttt{play} and \texttt{uber} contain dialogs for PlayStation and Uber, respectively.
We follow prior work~\cite{feigenblat_2021_tweetsumm} to preprocess the dialogs, keeping only those that contain 6--20 utterances between a single customer and agent.

We pose three types of aggregation queries over the six datasets via AISQL's \texttt{AI\_AGG()} function~\cite{liskowski_2025_cortex_aisql} with Llama 3.3 70B~\cite{grattafiori_2024_llama3}.
We issue separate summarization queries to \texttt{johns}, \texttt{village}, \texttt{play}, and \texttt{uber} (e.g., ``Summarize what customers are saying about John's Roast Pork'').
For \texttt{mcdonalds} and \texttt{airlines}, we issue comparative queries (e.g., ``Compare the reviews for the different McDonald's locations; highlight the commonalities and differences between the restaurant locations'') and ranking queries (e.g., ``Analyze the customer support dialogs for the different airline companies and describe the worst 3 airline companies regarding flight booking issues'').
We use the approach described in Section~\ref{sec:claims} to decompose each aggregate response into a list of claims.

To curate a diverse and balanced benchmark, we start from these automatically decomposed claims.
We use some directly, adapt others (e.g., recasting a reported proportion as a count, or negating a claim to obtain a false variant), and add manually written claims.
In total, we obtain 32 claims that span all claim types described in Section~\ref{sec:claims}, with an even split between grounded and ungrounded claims.
Table~\ref{tab:claims} shows a representative sample.
In general, complete human annotation of claim-level ground truth is infeasible in our setting because verifying quantified claims (e.g., ``all reviews are positive'') may require reviewing every tuple in the relation---over 1,000 per dataset.
Instead, the ground truth validity of each claim is determined with a reference implementation of \system\ that executes manually written verification queries.
Each invocation of a semantic operator takes the majority vote from an ensemble of three strong LLMs spanning distinct families (Claude Opus 4.8~\cite{anthropic_2026_claude_opus_4_8}, GPT-5.5~\cite{openai_2026_gpt_5_5}, and Gemini 3.5 Flash~\cite{gemini_3_5_flash}).
Moreover, the reference implementation disables all optimizations, since some can degrade quality and early stopping prevents capturing tuple-level labels for provenance and semantic operator quality evaluations.
To evaluate agreement of the reference labels against an independent human reference, we first draw a stratified random sample of per-operator decisions (i.e., the ensemble's tuple-level filter and map outputs) covering all 26 distinct semantic operators in our benchmark.
We then sample four decisions per operator, balanced across label values, for a total of 104 decisions.
The first author, blind to the ensemble labels, annotates each sampled decision, and we measure agreement between the human and ensemble using exact match and Cohen's $\kappa$.
The human and ensemble agree on 98.1\% of decisions (95\% Wilson CI $[0.93, 0.99]$; $\kappa = 0.96$). 
As an additional check, for claims whose verdict can be determined via a handful of tuples (e.g., $\exists$ or $\exists_{\ge k}$ with small $k$), we manually confirmed the ensemble's verdict: these checks agreed with the ensemble in every case.

\noindent
\textbf{Implementations.}
We use Snowflake Cortex AI~\cite{snowflake_cortex_ai} for LLM inference and embedding generation.
All embeddings are created by Snowflake arctic-embed-l-v2.0~\cite{merrick_2024_snowflake_arctic_embed}.
We compare \system\ with four implementations: \baserm, \ragagent, \rlm, and \evgunopt.

\baserm\ prompts an LLM to reason about whether a claim is supported by the given input relation and represents existing LLM-as-a-judge baselines~\cite{min_2023_factscore,tang_2024_minicheck,zheng_2023_llm_judge}.
We instantiate \baserm\ with three LLMs: Claude Opus 4.6~\cite{anthropic_2026_claude_opus_4_6}, Sonnet 4.6~\cite{anthropic_2026_claude_sonnet_4_6}, and Haiku 4.5~\cite{anthropic_2026_claude_haiku_4_5}.
It serializes each tuple as JSON and greedily fills the prompt up to Claude's 200k-token context window.
When claims involve vague quantifiers (e.g., ``common''), we provide hints specifying exact thresholds (e.g., ``at least 10\%'') for a fairer comparison with the reference ground truth.
These hints are also used by the other implementations.
We set temperature to 0 and allow up to 10 retries, since \baserm\ occasionally exhausts its 8,192-token output limit.

\ragagent\ is an LLM agent that uses a retrieval tool and represents an implementation of iterative retrieval-based claim verification approaches~\cite{khattab_2021_baleen,xie_2025_fire}.
The interface to the retrieval tool is similar to relevance sorting but restricted to top-k queries.
In addition to a semantic search query, inclusion keywords, and exclusion keywords, the agent must specify a limit on the most relevant tuples to retrieve from the database.
The agent can also include structured filters in the retrieval query (e.g., \texttt{business\_id = [A]}), which are executed as pre-filters before retrieving the requested number of tuples.
After tuples are retrieved, the LLM can choose the number of tuples to view initially from the set $\{10, 20, \ldots, 50\}$.
At each iteration of the agentic loop, the LLM either (1) issues a retrieval query and views the initial results, (2) views the next page of previously retrieved tuples, or (3) outputs whether the claim is grounded and terminates.
We instantiate \ragagent\ with the same three Claude models as \baserm\ and set temperature to 1.

\rlm\ is a Recursive Language Model (RLM)~\cite{zhang_2026_rlm} that can iteratively write and execute arbitrary Python code in a REPL, where the generated code can also invoke sub-LLM calls.
We additionally provide \rlm\ with the same retrieval tool given to \ragagent.
As such, this implementation is a generalization of prior imperative, program-based verification approaches~\cite{pan_2023_programfc,hu_2025_boost}.
We instantiate \rlm\ using the recent implementation in DSPy~\cite{khattab_2024_dspy} and keep the root LLM fixed to Opus while varying the sub-LLM to the same three Claude models used for \baserm\ and \ragagent.
The root LLM uses a temperature of 1 while the sub-LLM uses a temperature of 0.

\evgunopt\ instantiates \system\ with no optimizations enabled and represents a naive semantic query processing engine like \textsc{Binder}~\cite{cheng_2023_binder}.
It uses each of the three Claude models from earlier and evaluates semantic operators on every tuple.
All \system\ implementations use a temperature of 0.
Moreover, implementations process tuples in batches of 32, trading up to a batch's worth of unnecessary LLM calls for significantly lower latency.

\evgopt\ instantiates \system\ with all optimizations enabled.
We evaluate five configurations, each using a different LLM for its semantic operators.
We include the three Claude models used in the other implementations and two additional Qwen models (Qwen3-VL-235B-A22B~\cite{bai_2026_qwen3_vl} and Qwen3-Next-80B-A3B~\cite{qwen_2025_qwen3_next}) to evaluate how quality and cost change with even weaker models.
Opus is always used for the optimizer's LLM to generate high-quality search queries.
To maintain high recall, we set the threshold for similarity filters to 0.15.
For estimation with CSs, we set $\alpha = \error = 0.05$.

For \system\ implementations, claims are compiled to queries by providing an LLM with the claim, the aggregation prompt, the dataset schema, hints for vague quantifiers, and \system's API documentation (with example queries from a separate movie review domain).
Averaged over the 32 claims and three trials each, it costs \$0.051 $\pm$ 0.001 and takes 9.8 $\pm$ 1.2s (mean $\pm$ SD) to compile a claim using Opus 4.6, with negligible variation across claim types.
We reviewed every generated query and found each equivalent to its manually written reference query.
We attribute this reliable compilation to \system's declarative, constrained query interface.

\begin{table*}[t]
  \centering
  \caption{Verification quality, average cost, and average latency across claims. We report the mean [min, max] over three trials.}
  \label{tab:verification_results}
  \scriptsize
  \begin{tabular}{llcccccc}
  \toprule
  Implementation & LLM & Precision & Recall & F1 Score & Accuracy & Cost (\$) & Latency (s) \\
  \midrule
    \multirow{3}{*}{\baserm}
      & Claude Opus 4.6 & 0.71 [0.70, 0.73] & 0.94 [0.88, 1.00] & 0.81 [0.78, 0.84] & 0.78 [0.75, 0.81] & 0.945 [0.942, 0.947] & 47 [39, 58] \\
      & Claude Sonnet 4.6 & 0.69 [0.67, 0.71] & 0.90 [0.88, 0.94] & 0.78 [0.76, 0.81] & 0.75 [0.72, 0.78] & 0.570 [0.568, 0.573] & 92 [76, 105] \\
      & Claude Haiku 4.5 & 0.61 [0.60, 0.62] & 0.96 [0.94, 1.00] & 0.74 [0.73, 0.76] & 0.67 [0.66, 0.69] & 0.188 [0.188, 0.188] & 17 [17, 17] \\
  \addlinespace
    \multirow{3}{*}{\ragagent}
      & Claude Opus 4.6 & 0.70 [0.63, 0.75] & 0.75 [0.75, 0.75] & 0.72 [0.69, 0.75] & 0.71 [0.66, 0.75] & 2.61 [2.46, 2.74] & 220 [216, 225] \\
      & Claude Sonnet 4.6 & 0.73 [0.67, 0.76] & 0.79 [0.75, 0.81] & 0.76 [0.71, 0.79] & 0.75 [0.69, 0.78] & 2.84 [2.57, 3.00] & 324 [294, 355] \\
      & Claude Haiku 4.5 & 0.61 [0.60, 0.62] & 0.88 [0.75, 1.00] & 0.72 [0.67, 0.76] & 0.66 [0.62, 0.69] & 0.272 [0.258, 0.287] & 38 [33, 41] \\
  \addlinespace
    \multirow{3}{*}{\rlm}
      & Claude Opus 4.6 & 0.83 [0.80, 0.87] & 0.79 [0.75, 0.81] & 0.81 [0.77, 0.84] & 0.81 [0.78, 0.84] & 1.89 [1.77, 2.02] & 260 [257, 261] \\
      & Claude Sonnet 4.6 & 0.89 [0.86, 0.92] & 0.79 [0.75, 0.88] & 0.83 [0.80, 0.88] & 0.84 [0.81, 0.88] & 1.29 [1.21, 1.32] & 199 [190, 212] \\
      & Claude Haiku 4.5 & 0.80 [0.73, 0.86] & 0.73 [0.69, 0.75] & 0.76 [0.71, 0.80] & 0.77 [0.72, 0.81] & 0.649 [0.616, 0.680] & 177 [163, 193] \\
  \addlinespace
    \multirow{3}{*}{\evgunopt}
      & Claude Opus 4.6 & 0.89 [0.89, 0.89] & 1.00 [1.00, 1.00] & 0.94 [0.94, 0.94] & 0.94 [0.94, 0.94] & 15.88 [15.88, 15.88] & 331 [319, 353] \\
      & Claude Sonnet 4.6 & 0.83 [0.83, 0.83] & 0.94 [0.94, 0.94] & 0.88 [0.88, 0.88] & 0.88 [0.88, 0.88] & 9.54 [9.54, 9.54] & 304 [290, 326] \\
      & Claude Haiku 4.5 & 0.83 [0.83, 0.83] & 0.94 [0.94, 0.94] & 0.88 [0.88, 0.88] & 0.88 [0.88, 0.88] & 3.23 [3.23, 3.23] & 354 [316, 400] \\
  \addlinespace
    \multirow{5}{*}{\evgopt}
      & Claude Opus 4.6 & 0.89 [0.89, 0.89] & 1.00 [1.00, 1.00] & 0.94 [0.94, 0.94] & 0.94 [0.94, 0.94] & 5.05 [4.86, 5.36] & 261 [193, 317] \\
      & Claude Sonnet 4.6 & 0.88 [0.88, 0.88] & 0.94 [0.94, 0.94] & 0.91 [0.91, 0.91] & 0.91 [0.91, 0.91] & 3.25 [3.17, 3.37] & 192 [169, 227] \\
      & Claude Haiku 4.5 & 0.83 [0.83, 0.83] & 0.94 [0.94, 0.94] & 0.88 [0.88, 0.88] & 0.88 [0.88, 0.88] & 1.01 [1.00, 1.01] & 133 [103, 179] \\
      & Qwen3-VL & 0.83 [0.82, 0.83] & 0.92 [0.88, 0.94] & 0.87 [0.85, 0.88] & 0.86 [0.84, 0.88] & 0.183 [0.174, 0.188] & 228 [216, 234] \\
      & Qwen3-Next & 0.82 [0.81, 0.82] & 0.83 [0.81, 0.88] & 0.82 [0.81, 0.85] & 0.82 [0.81, 0.84] & 0.113 [0.111, 0.116] & 114 [111, 115] \\
  \bottomrule
\end{tabular}

\end{table*}

\begin{figure*}[t]
  \centering
  \includegraphics[width=\textwidth]{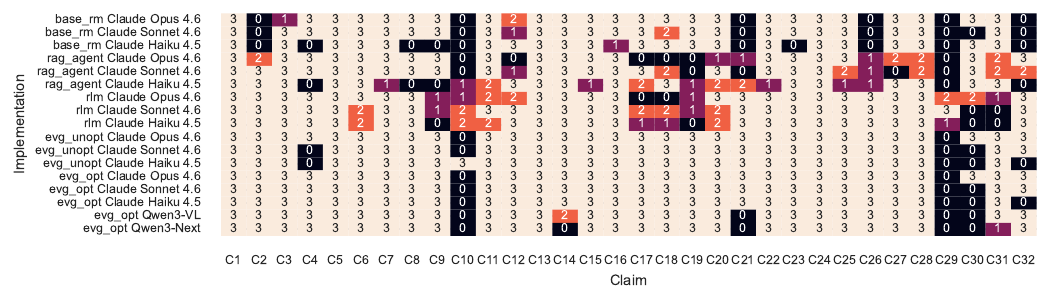}
  \caption{Per-claim verification correctness. Each cell shows the number of correct trials out of three.}
  \Description{Heatmap of per-claim verification correctness across implementations and models.}
  \label{fig:verification_result}
\end{figure*}

\noindent
\textbf{Verification Results.}
Figure~\ref{fig:cost_f1_pareto} highlights the cost--quality tradeoffs for each implementation, and Table~\ref{tab:verification_results} reports their quality, cost, and latency in more detail.
Precision, recall, and F1 score treat ungrounded claims as the positive class, measuring an implementation's ability to detect hallucinations.
Since our benchmark has an even split between grounded and ungrounded claims, we also report accuracy.
Cost is computed from input and output token counts at published per-million-token USD rates for each LLM.
We use Anthropic's input/output rates~\cite{claude_pricing} of \$5/\$25 for Opus 4.6, \$3/\$15 for Sonnet 4.6, and \$1/\$5 for Haiku 4.5, and Alibaba Cloud's rates~\cite{alibaba_cloud_pricing} of \$0.287/\$1.147 for Qwen3-VL and \$0.144/\$0.574 for Qwen3-Next.
Latency is wall-clock time but is subject to Cortex's server-side variability, so we treat it as a secondary metric.
Reported costs and latencies include claim compilation, optimization, and execution.

\evgopt\ configurations form the entire Pareto frontier: all other implementations are dominated by some \evgopt\ point with higher or equal F1 at lower cost.
Compared to \evgunopt\ with Opus, \evgopt\ preserves verification quality at an F1 of 0.94 while reducing cost by 3.1$\times$ (\$5.05 vs.\ \$15.88), demonstrating the effectiveness of \system's optimizations.
\rlm\ with Sonnet performs the best among the non-\system\ implementations with an F1 of 0.83; however, \evgopt\ with Qwen3-VL dominates at 0.87 with 7.0$\times$ lower cost (\$0.183 vs.\ \$1.29).
These results highlight the advantages of \system's declarative, neuro-symbolic design.
By enabling the LLM compiler to simply specify the high-level verification query, \system\ abstracts away optimization details and restricts LLMs to fine-grained semantic tasks.
This yields higher verification quality at lower cost with weaker models than approaches that rely on strong LLMs for both semantic and symbolic processing or program-based approaches that push optimization onto the LLM.

Figure~\ref{fig:verification_result} reveals each implementation's failure modes.
\baserm\ fails when grounding witnesses fall beyond its context window (e.g., a review praising the restaurant's chicken salad for existential claim C2, the target entities for ordinal claims C10 and C26, and all but one airline for nested claim C21).
\baserm\ also struggles with precise quantities, defaulting to ungrounded on cardinal claim C29 and proportional claim C32.
\ragagent\ avoids the first problem since retrieval surfaces the relevant witnesses, but because it estimates quantities in natural language from retrieved samples, it still fails on claims requiring precise aggregates (e.g., C10, C19, C26, and C29).
While \rlm\ improves over \ragagent, it sometimes processes only a sample of the relation with its sub-LLM, so it faces similar sampling imprecision (e.g., cardinal claim C30 and proportional claim C31).
Even when \rlm\ processes every tuple, noisy per-tuple labels and small per-group denominators flip the ranking for ordinal claim C9.
In contrast, \system's only source of error is semantic operator mislabeling, as its symbolic aggregation is otherwise exact or statistically sound.
Errors occur more frequently with weaker models; for example, on C30, only Opus returns the correct verdict.
However, claims involving close rankings (C10) or exact counts (C29) remain challenging even for the strongest model.

\begin{table*}[t]
  \centering
  \caption{\system's provenance precision and semantic operator quality. We report the mean [min, max] over three trials.}
  \label{tab:component_quality}
  \scriptsize
  \begin{tabular}{llccccc}
  \toprule
  Implementation & LLM & Provenance Precision & Filter Precision & Filter Recall & Filter F1 Score & Map Accuracy \\
  \midrule
    \multirow{3}{*}{\evgunopt}
      & Claude Opus 4.6 & 0.94 [0.94, 0.94] & 0.98 [0.98, 0.98] & 0.97 [0.97, 0.98] & 0.98 [0.98, 0.98] & 0.96 [0.96, 0.96] \\
      & Claude Sonnet 4.6 & 0.91 [0.91, 0.91] & 0.98 [0.98, 0.98] & 0.97 [0.97, 0.97] & 0.97 [0.97, 0.97] & 0.95 [0.95, 0.95] \\
      & Claude Haiku 4.5 & 0.83 [0.83, 0.83] & 0.93 [0.92, 0.93] & 0.99 [0.99, 0.99] & 0.96 [0.95, 0.96] & 0.93 [0.93, 0.93] \\
  \addlinespace
    \multirow{5}{*}{\evgopt}
      & Claude Opus 4.6 & 0.94 [0.94, 0.95] & 0.99 [0.98, 0.99] & 0.90 [0.89, 0.91] & 0.94 [0.93, 0.95] & 0.94 [0.94, 0.95] \\
      & Claude Sonnet 4.6 & 0.89 [0.88, 0.89] & 0.97 [0.97, 0.98] & 0.89 [0.88, 0.90] & 0.93 [0.93, 0.94] & 0.92 [0.92, 0.92] \\
      & Claude Haiku 4.5 & 0.82 [0.81, 0.82] & 0.95 [0.94, 0.97] & 0.89 [0.89, 0.90] & 0.92 [0.92, 0.93] & 0.89 [0.88, 0.89] \\
      & Qwen3-VL & 0.89 [0.89, 0.89] & 0.98 [0.98, 0.98] & 0.87 [0.84, 0.89] & 0.92 [0.91, 0.94] & 0.92 [0.91, 0.93] \\
      & Qwen3-Next & 0.86 [0.86, 0.86] & 0.98 [0.98, 0.98] & 0.83 [0.83, 0.84] & 0.90 [0.90, 0.91] & 0.88 [0.87, 0.88] \\
  \bottomrule
\end{tabular}

\end{table*}

\noindent
\textbf{Provenance and Semantic Operator Quality.}
To understand how \system's provenance and semantic operators degrade with weaker models, we execute the reference queries with \evgunopt\ and \evgopt\ across LLMs and measure provenance precision and per-operator quality (Table~\ref{tab:component_quality}).
Semantic filter and map quality are micro-averaged over distinct semantic operators, while provenance precision is pooled across all queries.
Provenance precision is the proportion of returned tokens that are correct: a positive token $\postoken_{\tuple}$ is correct if $\formula(\tuple) = \top$ according to the reference, and dually for a negative token $\negtoken_{\tuple}$.
We report precision rather than recall because \system\ intentionally returns non-exhaustive provenance tokens; we therefore require provenance to be correct, not complete.
We measure filter quality by precision, recall, and F1, treating tuples selected by the reference's filter as the positive class.
Map accuracy is the proportion of map outputs that match the reference.
\evgopt\ with Opus closely tracks \evgunopt: it matches provenance precision at 0.94 and has comparable filter precision (0.99 vs.\ 0.98).
Similarity filtering modestly reduces filter recall to 0.90 from 0.97, lowering filter F1 to 0.94 from 0.98, while map accuracy dips slightly to 0.94 from 0.96.
Quality then degrades gracefully as the model weakens: filter F1 and map accuracy remain at least 0.90 and 0.88, respectively, across all models, while provenance precision is the most sensitive, dropping to 0.82 from 0.94.

\begin{figure*}[t]
  \centering
  \begin{subfigure}[t]{0.32\textwidth}
    \includegraphics{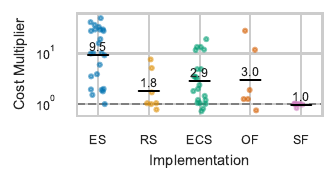}
    \caption{Optimization cost multipliers.}
    \label{fig:ablation}
  \end{subfigure}\hfill
  \begin{subfigure}[t]{0.32\textwidth}
    \includegraphics{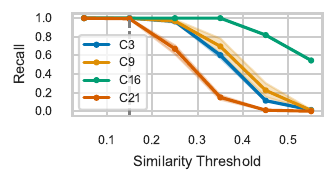}
    \caption{Similarity filter recall.}
    \label{fig:sim_filter_recall}
  \end{subfigure}\hfill
  \begin{subfigure}[t]{0.32\textwidth}
    \includegraphics{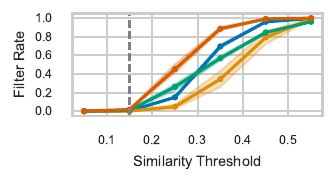}
    \caption{Similarity filter rate.}
    \label{fig:sim_filter_rate}
  \end{subfigure}
  \caption{Optimization analysis. (a) Cost multipliers (log scale) when each optimization is disabled, relative to \evgopt\ with Haiku 4.5 (ES: Early Stopping, RS: Relevance Sorting, ECS: Estimation with Confidence Sequences, OF: Operator Fusion, SF: Similarity Filtering). Each point is one claim's multiplier (mean of three trials); the horizontal line is the geometric mean and the dashed line at $1\times$ marks parity with \evgopt. (b) Recall and (c) filter rate of the similarity filter across thresholds; the dashed line marks the 0.15 default. Points are means over three trials, and shaded regions show the range.}
  \label{fig:optimization_analysis}
  \Description{Optimization analysis.}
\end{figure*}

\noindent
\textbf{Ablation Study.}
We evaluate the contribution of each optimization by performing a leave-one-out ablation study with Haiku, disabling one optimization while keeping the others enabled.
Figure~\ref{fig:ablation} reports the resulting cost multipliers relative to the fully optimized configuration; we summarize each optimization by the geometric mean of these ratios over its applicable claims.
Note that disabling early stopping effectively disables relevance sorting and estimation.

Early stopping leads to the largest cost savings of 9.5$\times$, illustrating the benefits of verification-aware optimizations.
Estimation and operator fusion each give comparable savings of 2.9$\times$ to 3.0$\times$.
Relevance sorting provides more moderate savings of 1.8$\times$.
Similarity filtering produces almost no cost savings in our evaluation because the similarity threshold is set low to maintain high recall, filtering out few tuples before the semantic filter is executed.
Prompt caching is omitted from Figure~\ref{fig:ablation}: four of its eight applicable claims are served entirely from cache and cost nothing when caching is enabled, so their cost multipliers (and hence the geometric mean) are undefined.
Measured instead by total cost across all eight claims, prompt caching reduces cost by 2.0$\times$, since duplicate LLM calls are served from cache rather than recomputed.

We examine the similarity filter's sensitivity to the threshold.
The benchmark's semantic filters reduce to four distinct predicates, those of C3, C9, C16, and C21; the rest reuse one of these on the same dataset and yield identical curves.
Figures~\ref{fig:sim_filter_recall} and \ref{fig:sim_filter_rate} report recall (the fraction of reference-selected tuples above the threshold) and filter rate (the fraction of tuples removed) across thresholds.
At the default of 0.15, recall exceeds 0.99 on all four filters with a filter rate of at most 0.02.
At higher thresholds, claims with \emph{semantically specific} filter predicates (e.g., C16: ``The restaurant review \{text\} mentions the number of whiskey varieties available'') maintain high recall, while claims with \emph{semantically broader} predicates (e.g., C21: ``The customer support \{dialog\} contains a customer complaint'') experience a rapid drop in recall.
Thus, 0.15 is a conservative default.
Integrating more advanced techniques (e.g., learning the threshold~\cite{shankar_2025_docetl,patel_2025_lotus} or late interaction~\cite{khattab_2020_colbert,santhanam_2022_colbertv2}) is a natural next step.

Four of the six optimizations---relevance sorting, operator fusion, similarity filtering, and prompt caching---maintain verification quality, while disabling early stopping or estimation each lowers precision by roughly 0.06.
Both drops trace to C4, a grounded universal claim.
With early stopping or estimation enabled, \system\ only processes a subset of tuples and correctly finds no counterexample.
Without these optimizations, processing all tuples increases the LLM's likelihood of flagging a false counterexample, flipping the verdict.
This effect is not general, however, since estimation can also miss a valid counterexample by sampling.
Overall, our optimizations yield substantial cost savings while maintaining quality.

\section{Related Work}
\label{sec:related}

Section~\ref{sec:introduction} presented work relevant to claim verification.
Here, we discuss other related work in more detail.

\textbf{Semantic query processing engines} such as Snowflake's Cortex AISQL~\cite{liskowski_2025_cortex_aisql}, DocETL~\cite{shankar_2025_docetl,wei_2026_moar}, LOTUS~\cite{patel_2025_lotus}, and Palimpzest~\cite{liu_2025_palimpzest,russo_2026_abacus}, among many others~\cite{urban_2023_caesura,lu_2025_vectraflow,anderson_2025_aryn,chen_2025_continuous_prompts,sanmartino_2026_stretto,santos_2025_samsara,li_2025_docdb,wang_2025_aop,cheng_2023_binder,databricks_ai_functions,duckdb_prompt,bigquery_ai_functions,dai_2024_uqe}, have emerged in recent years.
At their core is a set of semantic operators---including semantic aggregation---which augment traditional database operators with LLMs.
To reduce the cost of invoking these operators, novel query optimizers~\cite{russo_2026_abacus,wei_2026_moar} and efficient execution strategies~\cite{sun_2025_quest,zeighami_2025_bargain,liu_2025_sem_op_kv,urban_2024_eleet,shankar_2026_task_cascades,zhao_2025_llm_order_by,zeighami_2025_fdj} have been proposed, with recent work introducing the SemBench benchmark for evaluating these systems~\cite{lao_2025_sembench}.
Yet, the community lacks a systematic approach to verify the correctness of semantic aggregates.
\citeauthor{lee_2025_sic}~\cite{lee_2025_sic} describe a vision for semantic integrity constraints as a declarative abstraction for enforcing the correctness of LLM outputs in semantic queries.
\system\ is a step towards this vision.

\textbf{Approximate query processing (AQP)} has a long history of trading accuracy for speed via sampling~\cite{hellerstein_1997_online_aggregation,chaudhuri_2017_aqp,agarwal_2013_blinkdb,park_2018_verdictdb,zhu_2025_pilotdb}.
Online aggregation~\cite{hellerstein_1997_online_aggregation} introduced progressive refinement of aggregate estimates, allowing users to stop early once estimates appear tight enough; however, continuously monitoring classical confidence intervals invalidates their error guarantees~\cite{johari_2017_peeking}.
\system's use of confidence sequences~\cite{darling_1967_cs,waudby_smith_2023_betting_cs} enables valid monitoring without a predetermined sample size.
More recent work explores approximating aggregation queries that involve evaluating multimodal semantic operators for filters (e.g., \textsc{ABae}~\cite{kang_2021_abae}, ThalamusDB~\cite{jo_2024_thalamusdb}) and the statistics of interest (e.g., InQuest~\cite{russo_2023_inquest}).
In contrast to these settings, which estimate bounds on the true aggregate values, verification queries only need to resolve boolean comparisons involving the aggregates.
This structure enables earlier termination: while traditional AQP relaxes a precise aggregate to an interval estimate, \system\ collapses the interval to a boolean verdict, needing only enough precision to resolve the comparison.

\textbf{Data provenance} originated from the database community~\cite{glavic_2021_data_provenance}, with foundational models including data lineage~\cite{cui_2000_lineage}, why- and where-provenance~\cite{buneman_2001_why_where}, and why-not provenance~\cite{chapman_2009_why_not}.
Seminal work by \citeauthor{green_2007_provenance_semirings}~\cite{green_2007_provenance_semirings} unifies several of these models under provenance semirings, also known as how-provenance.
Subsequent work extends semiring semantics to other settings, such as aggregation~\cite{amsterdamer_2011_aggregate_provenance} and full first-order logic with negation~\cite{gradel_2024_fol_provenance}.
Several database systems support provenance capture and querying, differing in provenance model or approach~\cite{glavic_2009_perm,psallidas_2018_smoke,lee_2019_pug,sen_2025_provsql,mohammed_2025_fade,mohammed_2023_smokedduck}; however, none support semantic queries or provenance for first-order logic.
To our knowledge, \system\ is the first to bring the two together, applying and extending semiring provenance to justify claim verdicts with a single minimal explanation.
In NLP, \citeauthor{zhang_2020_provenance_for_claims}~\cite{zhang_2020_provenance_for_claims,zhang_2021_inferring_provenance} formalize provenance for claims as a labeled graph over dependent web sources, tracing how a claim originated and evolved.
Unlike this setting, \system\ verifies quantified claims over independent tuples in a relation, motivating its choice of provenance semantics.

\section{Conclusions}
\label{sec:conclusion}

We present \system, a system that reliably and efficiently verifies claims extracted from semantic aggregates by compiling them into declarative semantic verification queries composed of standard relational and semantic operators.
This formulation admits a suite of complementary optimizations that substantially reduce cost while preserving verification quality.
Furthermore, each verdict is accompanied by a minimal explanation based on semiring provenance for first-order logic.
Looking ahead, a natural extension is to exploit shared computation across the many claims that arise from a semantic aggregate, akin to multi-query optimization in databases.
As semantic aggregation becomes a standard feature of production data systems, \system\ demonstrates that the declarative query abstractions already present in these systems can be efficiently leveraged to make their outputs reliable and explainable.

\begin{acks}
We thank the Snowflake Cortex AISQL team for their valuable support and feedback.
This material is based upon work supported by the National Science Foundation Graduate Research Fellowship Program under Grant Nos 2439559 and 2040433. Any opinions, findings, and conclusions or recommendations expressed in this material are those of the authors and do not necessarily reflect the views of the National Science Foundation. 
\end{acks} 

\balance

\bibliographystyle{ACM-Reference-Format}
\bibliography{main}

@misc{yelp_open_dataset,
  title        = {Yelp Open Dataset},
  author       = {Yelp},
  year         = {2023},
  howpublished = {\url{https://business.yelp.com/data/resources/open-dataset/}},
  note         = {Accessed: 2026-03-31}
}

@misc{customer_support_on_twitter,
	title={Customer Support on Twitter},
	url={https://www.kaggle.com/dsv/8841},
	DOI={10.34740/KAGGLE/DSV/8841},
	publisher={Kaggle},
	author={Stuart Axelbrooke},
	year={2017}
}

@inproceedings{feigenblat_2021_tweetsumm,
  title = "{TWEETSUMM} - A Dialog Summarization Dataset for Customer Service",
  author = "Feigenblat, Guy  and
    Gunasekara, Chulaka  and
    Sznajder, Benjamin  and
    Joshi, Sachindra  and
    Konopnicki, David  and
    Aharonov, Ranit",
  editor = "Moens, Marie-Francine  and
    Huang, Xuanjing  and
    Specia, Lucia  and
    Yih, Scott Wen-tau",
  booktitle = "Findings of the Association for Computational Linguistics: EMNLP 2021",
  month = nov,
  year = "2021",
  address = "Punta Cana, Dominican Republic",
  publisher = "Association for Computational Linguistics",
  url = "https://aclanthology.org/2021.findings-emnlp.24/",
  doi = "10.18653/v1/2021.findings-emnlp.24",
  pages = "245--260",
}

@misc{snowflake_cortex_ai,
  title        = {Snowflake Cortex AI},
  author       = {Snowflake},
  year         = {n.d.},
  howpublished = {\url{https://www.snowflake.com/en/product/features/cortex/}},
  note         = {Accessed: 2026-03-31}
}

@misc{databricks_ai_functions,
  title        = {Introducing AI Functions: Integrating Large Language Models with Databricks SQL},
  author       = {Patrick Wendell and Eric Peter and Nicolas Pelaez and Jianwei Xie and Vinny Vijeyakumaar and Linhong Liu and Shitao Li},
  year         = {2023},
  howpublished = {\url{https://www.databricks.com/blog/2023/04/18/introducing-ai-functions-integrating-large-language-models-databricks-sql.html}},
  note         = {Accessed: 2026-03-31}
}

@misc{duckdb_prompt,
  title        = {INTRODUCING THE PROMPT() FUNCTION: USE THE POWER OF LLMS WITH SQL!},
  author       = {Till Döhmen},
  year         = {2024},
  howpublished = {\url{https://motherduck.com/blog/sql-llm-prompt-function-gpt-models/}},
  note         = {Accessed: 2026-03-31}
}

@misc{bigquery_ai_functions,
  title        = {Announcing BigQuery-managed AI functions for better SQL},
  author       = {Jian He and Vaibhav Sethi},
  year         = {2025},
  howpublished = {\url{https://cloud.google.com/blog/products/data-analytics/sql-reimagined-for-the-ai-era-with-bigquery-ai-functions}},
  note         = {Accessed: 2026-03-31}
}

@article{patel_2025_lotus,
  author = {Patel, Liana and Jha, Siddharth and Pan, Melissa and Gupta, Harshit and Asawa, Parth and Guestrin, Carlos and Zaharia, Matei},
  title = {Semantic Operators and Their Optimization: Enabling LLM-Based Data Processing with Accuracy Guarantees in LOTUS},
  year = {2025},
  issue_date = {July 2025},
  publisher = {VLDB Endowment},
  volume = {18},
  number = {11},
  issn = {2150-8097},
  url = {https://doi.org/10.14778/3749646.3749685},
  doi = {10.14778/3749646.3749685},
  journal = {Proc. VLDB Endow.},
  month = jul,
  pages = {4171-4184},
  numpages = {14}
}

@article{shankar_2025_docetl,
  author = {Shankar, Shreya and Chambers, Tristan and Shah, Tarak and Parameswaran, Aditya G. and Wu, Eugene},
  title = {DocETL: Agentic Query Rewriting and Evaluation for Complex Document Processing},
  year = {2025},
  issue_date = {May 2025},
  publisher = {VLDB Endowment},
  volume = {18},
  number = {9},
  issn = {2150-8097},
  url = {https://doi.org/10.14778/3746405.3746426},
  doi = {10.14778/3746405.3746426},
  journal = {Proc. VLDB Endow.},
  month = may,
  pages = {3035-3048},
  numpages = {14}
}

@inproceedings{lu_2025_vectraflow,
  title = {VectraFlow: Integrating Vectors into Stream Processing},
  author = {Lu, Duo and Feng, Siming and Zhou, Jonathan and Solleza, Franco and Schwarzkopf, Malte and {\c{C}}etintemel, U{\u{g}}ur},
  year = {2025},
  booktitle = {CIDR},
}

@inproceedings{anderson_2025_aryn,
  title = {The Design of an LLM-powered Unstructured Analytics System},
  author = {Anderson, Eric and Fritz, Jonathan and Lee, Austin and Li, Bohou and Lindblad, Mark and Lindeman, Henry and Meyer, Alex and Parmar, Parth and Ranade, Tanvi and Shah, Mehul A. and others},
  year = {2025},
  booktitle = {CIDR},
}

@inproceedings{liu_2025_palimpzest,
  title = {Palimpzest: Optimizing AI-Powered Analytics with Declarative Query Processing},
  author = {Liu, Chunwei and Russo, Matthew and Cafarella, Michael and Cao, Lei and Chen, Peter Baille and Chen, Zui and Franklin, Michael and Kraska, Tim and Madden, Samuel and Shahout, Rana and Vitagliano, Gerardo},
  year = {2025},
  booktitle = {CIDR},
}

@inproceedings{wang_2025_aop,
  title = {AOP: Automated and Interactive LLM Pipeline Orchestration
for Answering Complex Queries},
  author = {Wang, Jiayi and Li, Guoliang},
  year = {2025},
  booktitle = {CIDR},
}

@inproceedings{urban_2023_caesura,
  title = {CAESURA: Language Models as Multi-Modal Query Planners},
  author = {Urban, Matthias and Binnig, Carsten},
  year = {2023},
  booktitle = {CIDR},
}

@misc{chen_2025_continuous_prompts,
  title={Continuous Prompts: LLM-Augmented Pipeline Processing over Unstructured Streams}, 
  author={Shu Chen and Deepti Raghavan and Uğur Çetintemel},
  year={2025},
  eprint={2512.03389},
  archivePrefix={arXiv},
  primaryClass={cs.DB},
  url={https://arxiv.org/abs/2512.03389}, 
}

@misc{sanmartino_2026_stretto,
  title={The Stretto Execution Engine for LLM-Augmented Data Systems}, 
  author={Gabriele Sanmartino and Matthias Urban and Paolo Papotti and Carsten Binnig},
  year={2026},
  eprint={2602.04430},
  archivePrefix={arXiv},
  primaryClass={cs.DB},
  url={https://arxiv.org/abs/2602.04430}, 
}

@misc{santos_2025_samsara,
  title={Towards a Multimodal Stream Processing System}, 
  author={Uélison Jean Lopes dos Santos and Alessandro Ferri and Szilard Nistor and Riccardo Tommasini and Carsten Binnig and Manisha Luthra},
  year={2025},
  eprint={2510.14631},
  archivePrefix={arXiv},
  primaryClass={cs.DB},
  url={https://arxiv.org/abs/2510.14631}, 
}

@article{li_2025_docdb,
  author = {Li, Zequn and Zhong, Yuanhao and Chai, Chengliang and Sun, Zhaoze and Deng, Yuhao and Yuan, Ye and Wang, Guoren and Cao, Lei},
  title = {DocDB: A Database for Unstructured Document Analysis},
  year = {2025},
  issue_date = {August 2025},
  publisher = {VLDB Endowment},
  volume = {18},
  number = {12},
  issn = {2150-8097},
  url = {https://doi.org/10.14778/3750601.3750678},
  doi = {10.14778/3750601.3750678},
  journal = {Proc. VLDB Endow.},
  month = aug,
  pages = {5387-5390},
  numpages = {4}
}

@article{sun_2025_quest,
  author = {Sun, Zhaoze and Chai, Chengliang and Deng, Qiyan and Jin, Kaisen and Guo, Xinyu and Han, Han and Yuan, Ye and Wang, Guoren and Cao, Lei},
  title = {QUEST: Query Optimization in Unstructured Document Analysis},
  year = {2025},
  issue_date = {July 2025},
  publisher = {VLDB Endowment},
  volume = {18},
  number = {11},
  issn = {2150-8097},
  url = {https://doi.org/10.14778/3749646.3749713},
  doi = {10.14778/3749646.3749713},
  journal = {Proc. VLDB Endow.},
  month = jul,
  pages = {4560-4573},
  numpages = {14}
}

@misc{russo_2026_abacus,
  title={Abacus: A Cost-Based Optimizer for Semantic Operator Systems}, 
  author={Matthew Russo and Chunwei Liu and Sivaprasad Sudhir and Gerardo Vitagliano and Michael Cafarella and Tim Kraska and Samuel Madden},
  year={2026},
  eprint={2505.14661},
  archivePrefix={arXiv},
  primaryClass={cs.DB},
  doi={https://doi.org/10.14778/3796195.3796215},
  url={https://arxiv.org/abs/2505.14661}, 
}

@misc{wei_2026_moar,
  title={Multi-Objective Agentic Rewrites for Unstructured Data Processing}, 
  author={Lindsey Linxi Wei and Shreya Shankar and Sepanta Zeighami and Yeounoh Chung and Fatma Ozcan and Aditya G. Parameswaran},
  year={2026},
  eprint={2512.02289},
  archivePrefix={arXiv},
  primaryClass={cs.DB},
  url={https://arxiv.org/abs/2512.02289}, 
}

@inproceedings{liu_2025_sem_op_kv,
  title={Optimizing {LLM} Queries in Relational Data Analytics Workloads},
  author={Shu Liu and Asim Biswal and Amog Kamsetty and Audrey Cheng and Luis Gaspar Schroeder and Liana Patel and Shiyi Cao and Xiangxi Mo and Ion Stoica and Joseph E. Gonzalez and Matei Zaharia},
  booktitle={Eighth Conference on Machine Learning and Systems},
  year={2025},
  url={https://openreview.net/forum?id=R7bK9yycHp}
}

@misc{shankar_2026_task_cascades,
  title={Task Cascades for Efficient Unstructured Data Processing}, 
  author={Shreya Shankar and Sepanta Zeighami and Aditya Parameswaran},
  year={2026},
  eprint={2601.05536},
  archivePrefix={arXiv},
  primaryClass={cs.DB},
  url={https://arxiv.org/abs/2601.05536}, 
}

@article{urban_2024_eleet,
  author = {Urban, Matthias and Binnig, Carsten},
  title = {ELEET: Efficient Learned Query Execution over Text and Tables},
  year = {2024},
  issue_date = {September 2024},
  publisher = {VLDB Endowment},
  volume = {17},
  number = {13},
  issn = {2150-8097},
  url = {https://doi.org/10.14778/3704965.3704989},
  doi = {10.14778/3704965.3704989},
  journal = {Proc. VLDB Endow.},
  month = sep,
  pages = {4867-4880},
  numpages = {14}
}

@misc{zhao_2025_llm_order_by,
  title={Access Paths for Efficient Ordering with Large Language Models}, 
  author={Fuheng Zhao and Jiayue Chen and Yiming Pan and Tahseen Rabbani and Sohaib and Divyakant Agrawal and Amr El Abbadi and Paritosh Aggarwal and Anupam Datta and Dimitris Tsirogiannis},
  year={2025},
  eprint={2509.00303},
  archivePrefix={arXiv},
  primaryClass={cs.DB},
  url={https://arxiv.org/abs/2509.00303}, 
}

@misc{zeighami_2025_fdj,
  title={Featurized-Decomposition Join: Low-Cost Semantic Joins with Guarantees}, 
  author={Sepanta Zeighami and Shreya Shankar and Aditya Parameswaran},
  year={2025},
  eprint={2512.05399},
  archivePrefix={arXiv},
  primaryClass={cs.DB},
  url={https://arxiv.org/abs/2512.05399}, 
}

@article{zeighami_2025_bargain,
  author = {Zeighami, Sepanta and Shankar, Shreya and Parameswaran, Aditya},
  title = {Cut Costs, Not Accuracy: LLM-Powered Data Processing with Guarantees},
  year = {2025},
  issue_date = {December 2025},
  publisher = {Association for Computing Machinery},
  address = {New York, NY, USA},
  volume = {3},
  number = {6},
  url = {https://doi.org/10.1145/3769776},
  doi = {10.1145/3769776},
  journal = {Proc. ACM Manag. Data},
  month = dec,
  articleno = {311},
  numpages = {26}
}

@inproceedings{dai_2024_uqe,
  title={{UQE}: A Query Engine for Unstructured Databases},
  author={Hanjun Dai and Bethany Yixin Wang and Xingchen Wan and Bo Dai and Sherry Yang and Azade Nova and Pengcheng Yin and Phitchaya Mangpo Phothilimthana and Charles Sutton and Dale Schuurmans},
  booktitle={The Thirty-eighth Annual Conference on Neural Information Processing Systems},
  year={2024},
  url={https://openreview.net/forum?id=t7SGOv5W5z}
}

@misc{liskowski_2025_cortex_aisql,
  title={Cortex AISQL: A Production SQL Engine for Unstructured Data}, 
  author={Paweł Liskowski and Benjamin Han and Paritosh Aggarwal and Bowei Chen and Boxin Jiang and Nitish Jindal and Zihan Li and Aaron Lin and Kyle Schmaus and Jay Tayade and Weicheng Zhao and Anupam Datta and Nathan Wiegand and Dimitris Tsirogiannis},
  year={2025},
  eprint={2511.07663},
  archivePrefix={arXiv},
  primaryClass={cs.DB},
  url={https://arxiv.org/abs/2511.07663}, 
}

@misc{lao_2025_sembench,
  title={SemBench: A Benchmark for Semantic Query Processing Engines}, 
  author={Jiale Lao and Andreas Zimmerer and Olga Ovcharenko and Tianji Cong and Matthew Russo and Gerardo Vitagliano and Michael Cochez and Fatma Özcan and Gautam Gupta and Thibaud Hottelier and H. V. Jagadish and Kris Kissel and Sebastian Schelter and Andreas Kipf and Immanuel Trummer},
  year={2025},
  eprint={2511.01716},
  archivePrefix={arXiv},
  primaryClass={cs.DB},
  url={https://arxiv.org/abs/2511.01716}, 
}

@article{lee_2025_sic,
author = {Lee, Alexander W. and Chan, Justin and Fu, Michael and Kim, Nicolas and Mehta, Akshay and Raghavan, Deepti and \c{C}etintemel, U\u{g}ur},
title = {Semantic Integrity Constraints: Declarative Guardrails for AI-Augmented Data Processing Systems},
year = {2025},
issue_date = {July 2025},
publisher = {VLDB Endowment},
volume = {18},
number = {11},
issn = {2150-8097},
url = {https://doi.org/10.14778/3749646.3749677},
doi = {10.14778/3749646.3749677},
journal = {Proc. VLDB Endow.},
month = jul,
pages = {4073-4080},
numpages = {8}
}

@article{madden_2003_tag,
  author = {Madden, Samuel and Franklin, Michael J. and Hellerstein, Joseph M. and Hong, Wei},
  title = {TAG: a Tiny AGgregation service for ad-hoc sensor networks},
  year = {2003},
  issue_date = {Winter 2002},
  publisher = {Association for Computing Machinery},
  address = {New York, NY, USA},
  volume = {36},
  number = {SI},
  issn = {0163-5980},
  url = {https://doi.org/10.1145/844128.844142},
  doi = {10.1145/844128.844142},
  journal = {SIGOPS Oper. Syst. Rev.},
  month = dec,
  pages = {131–146},
  numpages = {16}
}

@misc{claude_pricing,
  title        = {Claude API Docs: Pricing},
  author       = {{Anthropic}},
  year         = {2026},
  howpublished = {\url{https://platform.claude.com/docs/en/about-claude/pricing}},
  note         = {Accessed: 2026-03-31}
}

@misc{alibaba_cloud_pricing,
  title        = {Alibaba Cloud Model Studio: Model inference pricing},
  author       = {{Alibaba Cloud}},
  year         = {2026},
  howpublished = {\url{https://www.alibabacloud.com/help/en/model-studio/model-pricing}},
  note         = {Accessed: 2026-06-23}
}

@misc{grattafiori_2024_llama3,
  title={The Llama 3 Herd of Models}, 
  author={Aaron Grattafiori and Abhimanyu Dubey and Abhinav Jauhri and others},
  year={2024},
  eprint={2407.21783},
  archivePrefix={arXiv},
  primaryClass={cs.AI},
  url={https://arxiv.org/abs/2407.21783}, 
}

@misc{bai_2026_qwen3_vl,
  title={Qwen3-VL Technical Report}, 
  author={Shuai Bai and Yuxuan Cai and Ruizhe Chen and Keqin Chen and Xionghui Chen and Zesen Cheng and Lianghao Deng and Wei Ding and Chang Gao and Chunjiang Ge and Wenbin Ge and Zhifang Guo and Qidong Huang and Jie Huang and Fei Huang and Binyuan Hui and Shutong Jiang and Zhaohai Li and Mingsheng Li and Mei Li and Kaixin Li and Zicheng Lin and Junyang Lin and Xuejing Liu and Jiawei Liu and Chenglong Liu and Yang Liu and Dayiheng Liu and Shixuan Liu and Dunjie Lu and Ruilin Luo and Chenxu Lv and Rui Men and Lingchen Meng and Xuancheng Ren and Xingzhang Ren and Sibo Song and Yuchong Sun and Jun Tang and Jianhong Tu and Jianqiang Wan and Peng Wang and Pengfei Wang and Qiuyue Wang and Yuxuan Wang and Tianbao Xie and Yiheng Xu and Haiyang Xu and Jin Xu and Zhibo Yang and Mingkun Yang and Jianxin Yang and An Yang and Bowen Yu and Fei Zhang and Hang Zhang and Xi Zhang and Bo Zheng and Humen Zhong and Jingren Zhou and Fan Zhou and Jing Zhou and Yuanzhi Zhu and Ke Zhu},
  year={2025},
  eprint={2511.21631},
  archivePrefix={arXiv},
  primaryClass={cs.CV},
  url={https://arxiv.org/abs/2511.21631}, 
}

@misc{qwen_2025_qwen3_next,
  title        = {Qwen3-Next: Towards Ultimate Training \& Inference Efficiency},
  author       = {{QwenTeam}},
  year         = {2025},
  howpublished = {\url{https://qwen.ai/blog?id=qwen3-next}},
  note         = {Accessed: 2026-06-23}
}

@misc{openai_2026_gpt_5_5,
  title        = {GPT-5.5 System Card},
  author       = {{OpenAI}},
  year         = {2026},
  howpublished = {\url{https://deploymentsafety.openai.com/gpt-5-5/gpt-5-5.pdf}},
  note         = {Accessed: 2026-06-23}
}

@misc{anthropic_2026_claude_opus_4_8,
  title        = {System Card: Claude Opus 4.8},
  author       = {{Anthropic}},
  year         = {2026},
  howpublished = {\url{https://www-cdn.anthropic.com/0b4915911bb0d19eca5b5ee635c80fef830a37ea.pdf}},
  note         = {Accessed: 2026-06-23}
}

@misc{anthropic_2026_claude_opus_4_6,
  title        = {System Card: Claude Opus 4.6},
  author       = {{Anthropic}},
  year         = {2026},
  howpublished = {\url{https://www-cdn.anthropic.com/0dd865075ad3132672ee0ab40b05a53f14cf5288.pdf}},
  note         = {Accessed: 2026-03-31}
}

@misc{anthropic_2026_claude_sonnet_4_6,
  title        = {System Card: Claude Sonnet 4.6},
  author       = {{Anthropic}},
  year         = {2026},
  howpublished = {\url{https://www-cdn.anthropic.com/78073f739564e986ff3e28522761a7a0b4484f84.pdf}},
  note         = {Accessed: 2026-03-31}
}

@misc{anthropic_2026_claude_haiku_4_5,
  title        = {System Card: Claude haiku 4.5},
  author       = {{Anthropic}},
  year         = {2025},
  howpublished = {\url{https://www-cdn.anthropic.com/7aad69bf12627d42234e01ee7c36305dc2f6a970.pdf}},
  note         = {Accessed: 2026-03-31}
}

@misc{gemini_3_5_flash,
  title        = {Gemini 3.5 Flash Model Card},
  author       = {{Google}},
  year         = {2026},
  howpublished = {\url{https://storage.googleapis.com/deepmind-media/Model-Cards/Gemini-3-5-Flash-Model-Card.pdf}},
  note         = {Accessed: 2026-06-23}
}

@misc{merrick_2024_snowflake_arctic_embed,
  title={Arctic-Embed: Scalable, Efficient, and Accurate Text Embedding Models}, 
  author={Luke Merrick and Danmei Xu and Gaurav Nuti and Daniel Campos},
  year={2024},
  eprint={2405.05374},
  archivePrefix={arXiv},
  primaryClass={cs.CL},
  url={https://arxiv.org/abs/2405.05374}, 
}

@inproceedings{cormack_2009_rrf,
  author = {Cormack, Gordon V. and Clarke, Charles L A and Buettcher, Stefan},
  title = {Reciprocal rank fusion outperforms condorcet and individual rank learning methods},
  year = {2009},
  isbn = {9781605584836},
  publisher = {Association for Computing Machinery},
  address = {New York, NY, USA},
  url = {https://doi.org/10.1145/1571941.1572114},
  doi = {10.1145/1571941.1572114},
  booktitle = {Proceedings of the 32nd International ACM SIGIR Conference on Research and Development in Information Retrieval},
  pages = {758–759},
  numpages = {2},
  location = {Boston, MA, USA},
  series = {SIGIR '09}
}

@inproceedings{chen_2020_tabfact,
  title={TabFact: A Large-scale Dataset for Table-based Fact Verification},
  author={Wenhu Chen and Hongmin Wang and Jianshu Chen and Yunkai Zhang and Hong Wang and Shiyang Li and Xiyou Zhou and William Yang Wang},
  booktitle={International Conference on Learning Representations},
  year={2020},
  url={https://openreview.net/forum?id=rkeJRhNYDH}
}

@inproceedings{wei_2024_safe,
  author = {Wei, Jerry and Yang, Chengrun and Song, Xinying and Lu, Yifeng and Hu, Nathan and Huang, Jie and Tran, Dustin and Peng, Daiyi and Liu, Ruibo and Huang, Da and Du, Cosmo and Le, Quoc V.},
  booktitle = {Advances in Neural Information Processing Systems},
  doi = {10.52202/079017-2567},
  editor = {A. Globerson and L. Mackey and D. Belgrave and A. Fan and U. Paquet and J. Tomczak and C. Zhang},
  pages = {80756--80827},
  publisher = {Curran Associates, Inc.},
  title = {Long-form factuality in large language models},
  url = {https://proceedings.neurips.cc/paper_files/paper/2024/file/937ae0e83eb08d2cb8627fe1def8c751-Paper-Conference.pdf},
  volume = {37},
  year = {2024}
}

@inproceedings{min_2023_factscore,
  title = "{FA}ct{S}core: Fine-grained Atomic Evaluation of Factual Precision in Long Form Text Generation",
  author = "Min, Sewon  and
    Krishna, Kalpesh  and
    Lyu, Xinxi  and
    Lewis, Mike  and
    Yih, Wen-tau  and
    Koh, Pang  and
    Iyyer, Mohit  and
    Zettlemoyer, Luke  and
    Hajishirzi, Hannaneh",
  editor = "Bouamor, Houda  and
    Pino, Juan  and
    Bali, Kalika",
  booktitle = "Proceedings of the 2023 Conference on Empirical Methods in Natural Language Processing",
  month = dec,
  year = "2023",
  address = "Singapore",
  publisher = "Association for Computational Linguistics",
  url = "https://aclanthology.org/2023.emnlp-main.741/",
  doi = "10.18653/v1/2023.emnlp-main.741",
  pages = "12076--12100",
}

@inproceedings{tang_2024_minicheck,
  title = "{M}ini{C}heck: Efficient Fact-Checking of {LLM}s on Grounding Documents",
  author = "Tang, Liyan  and
    Laban, Philippe  and
    Durrett, Greg",
  editor = "Al-Onaizan, Yaser  and
    Bansal, Mohit  and
    Chen, Yun-Nung",
  booktitle = "Proceedings of the 2024 Conference on Empirical Methods in Natural Language Processing",
  month = nov,
  year = "2024",
  address = "Miami, Florida, USA",
  publisher = "Association for Computational Linguistics",
  url = "https://aclanthology.org/2024.emnlp-main.499/",
  doi = "10.18653/v1/2024.emnlp-main.499",
  pages = "8818--8847",
}

@inproceedings{zheng_2023_llm_judge,
  author = {Zheng, Lianmin and Chiang, Wei-Lin and Sheng, Ying and Zhuang, Siyuan and Wu, Zhanghao and Zhuang, Yonghao and Lin, Zi and Li, Zhuohan and Li, Dacheng and Xing, Eric P. and Zhang, Hao and Gonzalez, Joseph E. and Stoica, Ion},
  title = {Judging LLM-as-a-judge with MT-bench and Chatbot Arena},
  year = {2023},
  publisher = {Curran Associates Inc.},
  address = {Red Hook, NY, USA},
  booktitle = {Proceedings of the 37th International Conference on Neural Information Processing Systems},
  articleno = {2020},
  numpages = {29},
  location = {New Orleans, LA, USA},
  series = {NIPS '23}
}

@inproceedings{xie_2025_fire,
title = "{FIRE}: Fact-checking with Iterative Retrieval and Verification",
author = "Xie, Zhuohan  and
  Xing, Rui  and
  Wang, Yuxia  and
  Geng, Jiahui  and
  Iqbal, Hasan  and
  Sahnan, Dhruv  and
  Gurevych, Iryna  and
  Nakov, Preslav",
editor = "Chiruzzo, Luis  and
  Ritter, Alan  and
  Wang, Lu",
booktitle = "Findings of the Association for Computational Linguistics: NAACL 2025",
month = apr,
year = "2025",
address = "Albuquerque, New Mexico",
publisher = "Association for Computational Linguistics",
url = "https://aclanthology.org/2025.findings-naacl.158/",
doi = "10.18653/v1/2025.findings-naacl.158",
pages = "2901--2914",
ISBN = "979-8-89176-195-7",
}

@inproceedings{khattab_2021_baleen,
 author = {Khattab, Omar and Potts, Christopher and Zaharia, Matei},
 booktitle = {Advances in Neural Information Processing Systems},
 editor = {M. Ranzato and A. Beygelzimer and Y. Dauphin and P.S. Liang and J. Wortman Vaughan},
 pages = {27670--27682},
 publisher = {Curran Associates, Inc.},
 title = {Baleen: Robust Multi-Hop Reasoning at Scale via Condensed Retrieval},
 url = {https://proceedings.neurips.cc/paper_files/paper/2021/file/e8b1cbd05f6e6a358a81dee52493dd06-Paper.pdf},
 volume = {34},
 year = {2021}
}

@inproceedings{khattab_2024_dspy,
  title={{DSP}y: Compiling Declarative Language Model Calls into State-of-the-Art Pipelines},
  author={Omar Khattab and Arnav Singhvi and Paridhi Maheshwari and Zhiyuan Zhang and Keshav Santhanam and Sri Vardhamanan A and Saiful Haq and Ashutosh Sharma and Thomas T. Joshi and Hanna Moazam and Heather Miller and Matei Zaharia and Christopher Potts},
  booktitle={The Twelfth International Conference on Learning Representations},
  year={2024},
  url={https://openreview.net/forum?id=sY5N0zY5Od}
}

@misc{zhang_2026_rlm,
  title={Recursive Language Models}, 
  author={Alex L. Zhang and Tim Kraska and Omar Khattab},
  year={2026},
  eprint={2512.24601},
  archivePrefix={arXiv},
  primaryClass={cs.AI},
  url={https://arxiv.org/abs/2512.24601}, 
}

@article{jayasekara_2025_cedar,
  author = {Jayasekara, Tharushi and Trummer, Immanuel},
  title = {CEDAR: A System for Cost-Efficient Data-Driven Claim Verification},
  year = {2025},
  issue_date = {July 2025},
  publisher = {VLDB Endowment},
  volume = {18},
  number = {11},
  issn = {2150-8097},
  url = {https://doi.org/10.14778/3749646.3749708},
  doi = {10.14778/3749646.3749708},
  journal = {Proc. VLDB Endow.},
  month = jul,
  pages = {4492-4504},
  numpages = {13}
}

@misc{theologitis_2026_thucy,
  title={Thucy: An LLM-based Multi-Agent System for Claim Verification across Relational Databases}, 
  author={Michael Theologitis and Dan Suciu},
  year={2026},
  eprint={2512.03278},
  archivePrefix={arXiv},
  primaryClass={cs.DB},
  url={https://arxiv.org/abs/2512.03278}, 
}

@article{karagiannis_2020_scrutinizer,
  author = {Karagiannis, Georgios and Saeed, Mohammed and Papotti, Paolo and Trummer, Immanuel},
  title = {Scrutinizer: a mixed-initiative approach to large-scale, data-driven claim verification},
  year = {2020},
  issue_date = {August 2020},
  publisher = {VLDB Endowment},
  volume = {13},
  number = {12},
  issn = {2150-8097},
  url = {https://doi.org/10.14778/3407790.3407841},
  doi = {10.14778/3407790.3407841},
  journal = {Proc. VLDB Endow.},
  month = jul,
  pages = {2508-2521},
  numpages = {14}
}

@inproceedings{jo_2019_aggchecker,
  author = {Jo, Saehan and Trummer, Immanuel and Yu, Weicheng and Wang, Xuezhi and Yu, Cong and Liu, Daniel and Mehta, Niyati},
  title = {Verifying Text Summaries of Relational Data Sets},
  year = {2019},
  isbn = {9781450356435},
  publisher = {Association for Computing Machinery},
  address = {New York, NY, USA},
  url = {https://doi.org/10.1145/3299869.3300074},
  doi = {10.1145/3299869.3300074},
  booktitle = {Proceedings of the 2019 International Conference on Management of Data},
  pages = {299-316},
  numpages = {18},
  location = {Amsterdam, Netherlands},
  series = {SIGMOD '19}
}

@misc{theologitis_2026_claimdb,
  title={ClaimDB: A Fact Verification Benchmark over Large Structured Data}, 
  author={Michael Theologitis and Preetam Prabhu Srikar Dammu and Chirag Shah and Dan Suciu},
  year={2026},
  eprint={2601.14698},
  archivePrefix={arXiv},
  primaryClass={cs.CL},
  url={https://arxiv.org/abs/2601.14698}, 
}

@inproceedings{chegini_2025_repanda,
  title = "{R}e{P}anda: Pandas-powered Tabular Verification and Reasoning",
  author = "Chegini, Atoosa  and
    Rezaei, Keivan  and
    Eghbalzadeh, Hamid  and
    Feizi, Soheil",
  editor = "Che, Wanxiang  and
    Nabende, Joyce  and
    Shutova, Ekaterina  and
    Pilehvar, Mohammad Taher",
  booktitle = "Proceedings of the 63rd Annual Meeting of the Association for Computational Linguistics (Volume 1: Long Papers)",
  month = jul,
  year = "2025",
  address = "Vienna, Austria",
  publisher = "Association for Computational Linguistics",
  url = "https://aclanthology.org/2025.acl-long.1549/",
  doi = "10.18653/v1/2025.acl-long.1549",
  pages = "32200--32212",
  ISBN = "979-8-89176-251-0",
}

@inproceedings{pan_2023_programfc,
  title = "Fact-Checking Complex Claims with Program-Guided Reasoning",
  author = "Pan, Liangming  and
    Wu, Xiaobao  and
    Lu, Xinyuan  and
    Luu, Anh Tuan  and
    Wang, William Yang  and
    Kan, Min-Yen  and
    Nakov, Preslav",
  editor = "Rogers, Anna  and
    Boyd-Graber, Jordan  and
    Okazaki, Naoaki",
  booktitle = "Proceedings of the 61st Annual Meeting of the Association for Computational Linguistics (Volume 1: Long Papers)",
  month = jul,
  year = "2023",
  address = "Toronto, Canada",
  publisher = "Association for Computational Linguistics",
  url = "https://aclanthology.org/2023.acl-long.386/",
  doi = "10.18653/v1/2023.acl-long.386",
  pages = "6981--7004",
}

@misc{hu_2025_boost,
  title={BOOST: Bootstrapping Strategy-Driven Reasoning Programs for Program-Guided Fact-Checking}, 
  author={Qisheng Hu and Quanyu Long and Wenya Wang},
  year={2025},
  eprint={2504.02467},
  archivePrefix={arXiv},
  primaryClass={cs.AI},
  url={https://arxiv.org/abs/2504.02467}, 
}

@inproceedings{cheng_2023_binder,
  title={Binding Language Models in Symbolic Languages},
  author={Zhoujun Cheng and Tianbao Xie and Peng Shi and Chengzu Li and Rahul Nadkarni and Yushi Hu and Caiming Xiong and Dragomir Radev and Mari Ostendorf and Luke Zettlemoyer and Noah A. Smith and Tao Yu},
  booktitle={The Eleventh International Conference on Learning Representations },
  year={2023},
  url={https://openreview.net/forum?id=lH1PV42cbF}
}

@inproceedings{wang_2023_folk,
  title = "Explainable Claim Verification via Knowledge-Grounded Reasoning with Large Language Models",
  author = "Wang, Haoran  and
    Shu, Kai",
  editor = "Bouamor, Houda  and
    Pino, Juan  and
    Bali, Kalika",
  booktitle = "Findings of the Association for Computational Linguistics: EMNLP 2023",
  month = dec,
  year = "2023",
  address = "Singapore",
  publisher = "Association for Computational Linguistics",
  url = "https://aclanthology.org/2023.findings-emnlp.416/",
  doi = "10.18653/v1/2023.findings-emnlp.416",
  pages = "6288--6304",
}

@article{hellerstein_1997_online_aggregation,
  author = {Hellerstein, Joseph M. and Haas, Peter J. and Wang, Helen J.},
  title = {Online aggregation},
  year = {1997},
  issue_date = {June 1997},
  publisher = {Association for Computing Machinery},
  address = {New York, NY, USA},
  volume = {26},
  number = {2},
  issn = {0163-5808},
  url = {https://doi.org/10.1145/253262.253291},
  doi = {10.1145/253262.253291},
  journal = {SIGMOD Rec.},
  month = jun,
  pages = {171-182},
  numpages = {12}
}

@inproceedings{chaudhuri_2017_aqp,
  author = {Chaudhuri, Surajit and Ding, Bolin and Kandula, Srikanth},
  title = {Approximate Query Processing: No Silver Bullet},
  year = {2017},
  isbn = {9781450341974},
  publisher = {Association for Computing Machinery},
  address = {New York, NY, USA},
  url = {https://doi.org/10.1145/3035918.3056097},
  doi = {10.1145/3035918.3056097},
  booktitle = {Proceedings of the 2017 ACM International Conference on Management of Data},
  pages = {511–519},
  numpages = {9},
  location = {Chicago, Illinois, USA},
  series = {SIGMOD '17}
}

@inproceedings{agarwal_2013_blinkdb,
  author = {Agarwal, Sameer and Mozafari, Barzan and Panda, Aurojit and Milner, Henry and Madden, Samuel and Stoica, Ion},
  title = {BlinkDB: queries with bounded errors and bounded response times on very large data},
  year = {2013},
  isbn = {9781450319942},
  publisher = {Association for Computing Machinery},
  address = {New York, NY, USA},
  url = {https://doi.org/10.1145/2465351.2465355},
  doi = {10.1145/2465351.2465355},
  booktitle = {Proceedings of the 8th ACM European Conference on Computer Systems},
  pages = {29-42},
  numpages = {14},
  location = {Prague, Czech Republic},
  series = {EuroSys '13}
}

@inproceedings{park_2018_verdictdb,
  author = {Park, Yongjoo and Mozafari, Barzan and Sorenson, Joseph and Wang, Junhao},
  title = {VerdictDB: Universalizing Approximate Query Processing},
  year = {2018},
  isbn = {9781450347037},
  publisher = {Association for Computing Machinery},
  address = {New York, NY, USA},
  url = {https://doi.org/10.1145/3183713.3196905},
  doi = {10.1145/3183713.3196905},
  booktitle = {Proceedings of the 2018 International Conference on Management of Data},
  pages = {1461-1476},
  numpages = {16},
  location = {Houston, TX, USA},
  series = {SIGMOD '18}
}

@article{zhu_2025_pilotdb,
author = {Zhu, Yuxuan and Jin, Tengjun and Baziotis, Stefanos and Zhang, Chengsong and Mendis, Charith and Kang, Daniel},
title = {PilotDB: Database-Agnostic Online Approximate Query Processing with A Priori Error Guarantees},
year = {2025},
issue_date = {June 2025},
publisher = {Association for Computing Machinery},
address = {New York, NY, USA},
volume = {3},
number = {3},
url = {https://doi.org/10.1145/3725335},
doi = {10.1145/3725335},
journal = {Proc. ACM Manag. Data},
month = jun,
articleno = {198},
numpages = {28}
}

@article{kang_2021_abae,
  author = {Kang, Daniel and Guibas, John and Bailis, Peter and Hashimoto, Tatsunori and Sun, Yi and Zaharia, Matei},
  title = {Accelerating approximate aggregation queries with expensive predicates},
  year = {2021},
  issue_date = {July 2021},
  publisher = {VLDB Endowment},
  volume = {14},
  number = {11},
  issn = {2150-8097},
  url = {https://doi.org/10.14778/3476249.3476285},
  doi = {10.14778/3476249.3476285},
  journal = {Proc. VLDB Endow.},
  month = jul,
  pages = {2341-2354},
  numpages = {14}
}

@article{russo_2023_inquest,
  author = {Russo, Matthew and Hashimoto, Tatsunori and Kang, Daniel and Sun, Yi and Zaharia, Matei},
  title = {Accelerating Aggregation Queries on Unstructured Streams of Data},
  year = {2023},
  issue_date = {July 2023},
  publisher = {VLDB Endowment},
  volume = {16},
  number = {11},
  issn = {2150-8097},
  url = {https://doi.org/10.14778/3611479.3611496},
  doi = {10.14778/3611479.3611496},
  journal = {Proc. VLDB Endow.},
  month = jul,
  pages = {2897-2910},
  numpages = {14}
}

@article{jo_2024_thalamusdb,
  author = {Jo, Saehan and Trummer, Immanuel},
  title = {ThalamusDB: Approximate Query Processing on Multi-Modal Data},
  year = {2024},
  issue_date = {June 2024},
  publisher = {Association for Computing Machinery},
  address = {New York, NY, USA},
  volume = {2},
  number = {3},
  url = {https://doi.org/10.1145/3654989},
  doi = {10.1145/3654989},
  journal = {Proc. ACM Manag. Data},
  month = may,
  articleno = {186},
  numpages = {26}
}

@article{waudby_smith_2023_betting_cs,
  author = {Waudby-Smith, Ian and Ramdas, Aaditya},
  title = {Estimating means of bounded random variables by betting},
  journal = {Journal of the Royal Statistical Society Series B: Statistical Methodology},
  volume = {86},
  number = {1},
  pages = {1-27},
  year = {2023},
  month = {02},
  issn = {1369-7412},
  doi = {10.1093/jrsssb/qkad009},
  url = {https://doi.org/10.1093/jrsssb/qkad009},
  eprint = {https://academic.oup.com/jrsssb/article-pdf/86/1/1/56961777/qkad009.pdf},
}

@article{darling_1967_cs,
  author = {D. A. Darling  and Herbert Robbins },
  title = {CONFIDENCE SEQUENCES FOR MEAN, VARIANCE, AND MEDIAN},
  journal = {Proceedings of the National Academy of Sciences},
  volume = {58},
  number = {1},
  pages = {66-68},
  year = {1967},
  doi = {10.1073/pnas.58.1.66},
  URL = {https://www.pnas.org/doi/abs/10.1073/pnas.58.1.66},
  eprint = {https://www.pnas.org/doi/pdf/10.1073/pnas.58.1.66}
}

@inproceedings{johari_2017_peeking,
  author = {Johari, Ramesh and Koomen, Pete and Pekelis, Leonid and Walsh, David},
  title = {Peeking at A/B Tests: Why it matters, and what to do about it},
  year = {2017},
  isbn = {9781450348874},
  publisher = {Association for Computing Machinery},
  address = {New York, NY, USA},
  url = {https://doi.org/10.1145/3097983.3097992},
  doi = {10.1145/3097983.3097992},
  booktitle = {Proceedings of the 23rd ACM SIGKDD International Conference on Knowledge Discovery and Data Mining},
  pages = {1517–1525},
  numpages = {9},
  location = {Halifax, NS, Canada},
  series = {KDD '17}
}

@article{cui_2000_lineage,
  author = {Cui, Yingwei and Widom, Jennifer and Wiener, Janet L.},
  title = {Tracing the lineage of view data in a warehousing environment},
  year = {2000},
  issue_date = {June 2000},
  publisher = {Association for Computing Machinery},
  address = {New York, NY, USA},
  volume = {25},
  number = {2},
  issn = {0362-5915},
  url = {https://doi.org/10.1145/357775.357777},
  doi = {10.1145/357775.357777},
  journal = {ACM Trans. Database Syst.},
  month = jun,
  pages = {179-227},
  numpages = {49}
}

@inproceedings{buneman_2001_why_where,
  author="Buneman, Peter
  and Khanna, Sanjeev
  and Wang-Chiew, Tan",
  editor="Van den Bussche, Jan
  and Vianu, Victor",
  title="Why and Where: A Characterization of Data Provenance",
  booktitle="Database Theory --- ICDT 2001",
  year="2001",
  publisher="Springer Berlin Heidelberg",
  address="Berlin, Heidelberg",
  pages="316--330",
  isbn="978-3-540-44503-6"
}

@article{glavic_2021_data_provenance,
  author = {Glavic, Boris},
  title = {Data Provenance},
  year = {2021},
  issue_date = {Apr 2021},
  publisher = {Now Publishers Inc.},
  address = {Hanover, MA, USA},
  volume = {9},
  number = {3-4},
  issn = {1931-7883},
  url = {https://doi.org/10.1561/1900000068},
  doi = {10.1561/1900000068},
  journal = {Found. Trends Databases},
  month = apr,
  pages = {209-441},
  numpages = {236}
}

@inproceedings{chapman_2009_why_not,
  author = {Chapman, Adriane and Jagadish, H. V.},
  title = {Why not?},
  year = {2009},
  isbn = {9781605585512},
  publisher = {Association for Computing Machinery},
  address = {New York, NY, USA},
  url = {https://doi.org/10.1145/1559845.1559901},
  doi = {10.1145/1559845.1559901},
  booktitle = {Proceedings of the 2009 ACM SIGMOD International Conference on Management of Data},
  pages = {523-534},
  numpages = {12},
  location = {Providence, Rhode Island, USA},
  series = {SIGMOD '09}
}

@inproceedings{green_2007_provenance_semirings,
  author = {Green, Todd J. and Karvounarakis, Grigoris and Tannen, Val},
  title = {Provenance semirings},
  year = {2007},
  isbn = {9781595936851},
  publisher = {Association for Computing Machinery},
  address = {New York, NY, USA},
  url = {https://doi.org/10.1145/1265530.1265535},
  doi = {10.1145/1265530.1265535},
  booktitle = {Proceedings of the Twenty-Sixth ACM SIGMOD-SIGACT-SIGART Symposium on Principles of Database Systems},
  pages = {31-40},
  numpages = {10},
  location = {Beijing, China},
  series = {PODS '07}
}

@inproceedings{amsterdamer_2011_aggregate_provenance,
  author = {Amsterdamer, Yael and Deutch, Daniel and Tannen, Val},
  title = {Provenance for aggregate queries},
  year = {2011},
  isbn = {9781450306607},
  publisher = {Association for Computing Machinery},
  address = {New York, NY, USA},
  url = {https://doi.org/10.1145/1989284.1989302},
  doi = {10.1145/1989284.1989302},
  booktitle = {Proceedings of the Thirtieth ACM SIGMOD-SIGACT-SIGART Symposium on Principles of Database Systems},
  pages = {153-164},
  numpages = {12},
  location = {Athens, Greece},
  series = {PODS '11}
}

@misc{gradel_2024_fol_provenance,
  title={Provenance Analysis and Semiring Semantics for First-Order Logic}, 
  author={Erich Grädel and Val Tannen},
  year={2024},
  eprint={2412.07986},
  archivePrefix={arXiv},
  primaryClass={cs.LO},
  url={https://arxiv.org/abs/2412.07986}, 
}

@article{psallidas_2018_smoke,
  author = {Psallidas, Fotis and Wu, Eugene},
  title = {Smoke: fine-grained lineage at interactive speed},
  year = {2018},
  issue_date = {February 2018},
  publisher = {VLDB Endowment},
  volume = {11},
  number = {6},
  issn = {2150-8097},
  url = {https://doi.org/10.14778/3199517.3199522},
  doi = {10.14778/3199517.3199522},
  journal = {Proc. VLDB Endow.},
  month = feb,
  pages = {719-732},
  numpages = {14}
}

@misc{sen_2025_provsql,
  title={ProvSQL: A General System for Keeping Track of the Provenance and Probability of Data}, 
  author={Aryak Sen and Silviu Maniu and Pierre Senellart},
  year={2025},
  eprint={2504.12058},
  archivePrefix={arXiv},
  primaryClass={cs.DB},
  url={https://arxiv.org/abs/2504.12058}, 
}

@inproceedings{mohammed_2023_smokedduck,
author = {Mohammed, Haneen and Summers, Charlie and Kaushik, Sughosh and Wu, Eugene},
title = {SmokedDuck Demonstration: SQLStepper},
year = {2023},
isbn = {9781450395076},
publisher = {Association for Computing Machinery},
address = {New York, NY, USA},
url = {https://doi.org/10.1145/3555041.3589731},
doi = {10.1145/3555041.3589731},
booktitle = {Companion of the 2023 International Conference on Management of Data},
pages = {183-186},
numpages = {4},
location = {Seattle, WA, USA},
series = {SIGMOD '23}
}

@article{mohammed_2025_fade,
  author = {Mohammed, Haneen and Yao, Alexander and Summers, Charlie and Zhong, Hongbin and Chan, Gromit Yeuk-Yin and Mitra, Subrata and Flokas, Lampros and Wu, Eugene},
  title = {FaDE: More Than a Million What-Ifs Per Second},
  year = {2024},
  issue_date = {December 2024},
  publisher = {VLDB Endowment},
  volume = {18},
  number = {4},
  issn = {2150-8097},
  url = {https://doi.org/10.14778/3717755.3717757},
  doi = {10.14778/3717755.3717757},
  journal = {Proc. VLDB Endow.},
  month = dec,
  pages = {943-955},
  numpages = {13}
}

@inproceedings{glavic_2009_perm,
  author = {Glavic, Boris and Alonso, Gustavo},
  title = {Perm: Processing Provenance and Data on the Same Data Model through Query Rewriting},
  year = {2009},
  isbn = {9780769535456},
  publisher = {IEEE Computer Society},
  address = {USA},
  url = {https://doi.org/10.1109/ICDE.2009.15},
  doi = {10.1109/ICDE.2009.15},
  booktitle = {Proceedings of the 2009 IEEE International Conference on Data Engineering},
  pages = {174–185},
  numpages = {12},
  series = {ICDE '09}
}

@article{lee_2019_pug,
  author = {Lee, Seokki and Lud\"{a}scher, Bertram and Glavic, Boris},
  title = {PUG: a framework and practical implementation for why and why-not provenance},
  year = {2019},
  issue_date = {February  2019},
  publisher = {Springer-Verlag},
  address = {Berlin, Heidelberg},
  volume = {28},
  number = {1},
  issn = {1066-8888},
  url = {https://doi.org/10.1007/s00778-018-0518-5},
  doi = {10.1007/s00778-018-0518-5},
  journal = {The VLDB Journal},
  month = feb,
  pages = {47–71},
  numpages = {25}
}

@inproceedings{zhang_2020_provenance_for_claims,
  title = "``Who said it, and Why?'' Provenance for Natural Language Claims",
  author = "Zhang, Yi  and
    Ives, Zachary  and
    Roth, Dan",
  editor = "Jurafsky, Dan  and
    Chai, Joyce  and
    Schluter, Natalie  and
    Tetreault, Joel",
  booktitle = "Proceedings of the 58th Annual Meeting of the Association for Computational Linguistics",
  month = jul,
  year = "2020",
  address = "Online",
  publisher = "Association for Computational Linguistics",
  url = "https://aclanthology.org/2020.acl-main.406/",
  doi = "10.18653/v1/2020.acl-main.406",
  pages = "4416--4426",
}

@inproceedings{zhang_2021_inferring_provenance,
  title = "What is Your Article Based On? Inferring Fine-grained Provenance",
  author = "Zhang, Yi  and
    Ives, Zachary  and
    Roth, Dan",
  editor = "Zong, Chengqing  and
    Xia, Fei  and
    Li, Wenjie  and
    Navigli, Roberto",
  booktitle = "Proceedings of the 59th Annual Meeting of the Association for Computational Linguistics and the 11th International Joint Conference on Natural Language Processing (Volume 1: Long Papers)",
  month = aug,
  year = "2021",
  address = "Online",
  publisher = "Association for Computational Linguistics",
  url = "https://aclanthology.org/2021.acl-long.458/",
  doi = "10.18653/v1/2021.acl-long.458",
}

@inproceedings{khattab_2020_colbert,
author = {Khattab, Omar and Zaharia, Matei},
title = {ColBERT: Efficient and Effective Passage Search via Contextualized Late Interaction over BERT},
year = {2020},
isbn = {9781450380164},
publisher = {Association for Computing Machinery},
address = {New York, NY, USA},
url = {https://doi.org/10.1145/3397271.3401075},
doi = {10.1145/3397271.3401075},
booktitle = {Proceedings of the 43rd International ACM SIGIR Conference on Research and Development in Information Retrieval},
pages = {39-48},
numpages = {10},
location = {Virtual Event, China},
series = {SIGIR '20}
}

@inproceedings{santhanam_2022_colbertv2,
  title = "{C}ol{BERT}v2: Effective and Efficient Retrieval via Lightweight Late Interaction",
  author = "Santhanam, Keshav  and
    Khattab, Omar  and
    Saad-Falcon, Jon  and
    Potts, Christopher  and
    Zaharia, Matei",
  editor = "Carpuat, Marine  and
    de Marneffe, Marie-Catherine  and
    Meza Ruiz, Ivan Vladimir",
  booktitle = "Proceedings of the 2022 Conference of the North American Chapter of the Association for Computational Linguistics: Human Language Technologies",
  month = jul,
  year = "2022",
  address = "Seattle, United States",
  publisher = "Association for Computational Linguistics",
  url = "https://aclanthology.org/2022.naacl-main.272/",
  doi = "10.18653/v1/2022.naacl-main.272",
  pages = "3715--3734",
}


\end{document}